\documentclass[twocolumn,showpacs,preprintnumbers,amsmath,amssymb,floatfix]{revtex4}
\usepackage{graphicx}

\usepackage[center]{subfigure}

\newcommand \rt {\right}
\newcommand \lt {\left}
\newcommand \be {\begin{equation}}
\newcommand \ee {\end{equation}}
\newcommand \ben {\begin{eqnarray}}
\newcommand \een {\end{eqnarray}}
\newcommand \nline {\nonumber \\}

\newcommand \dr {\delta\rho}

\newcommand \liqA {\rho_{\ell}^A}
\newcommand \liqB {\rho_{\ell}^B}
\newcommand \liqi {\rho_{\ell}^i}
\newcommand \liqr {\rho_{\ell}}
\newcommand \rhoA {\rho_A}
\newcommand \rhoB {\rho_B}
\newcommand \rhoi {\rho_i}

\newcommand \drhoA {\delta\rho_A}
\newcommand \drhoB {\delta\rho_B}
\newcommand \drhoi {\delta\rho_i}
\newcommand \drhoj {\delta\rho_j}
\newcommand \drho {\delta\rho}
\newcommand \dc {\delta c}
\newcommand \dN {\delta N}

\newcommand \cAA[1] {\hat{C}_{#1}^{AA}}
\newcommand \cBB[1] {\hat{C}_{#1}^{BB}}
\newcommand \cAB[1] {\hat{C}_{#1}^{AB}}
\newcommand \cij[1] {\hat{C}_{#1}^{ij}}
\newcommand \cb[1] {\hat{\bar{C}}_{#1}}

\newcommand \Chat[1] {\hat{C}_{#1}}
\newcommand \Chatm {\hat{C}_{m}}
\newcommand \dChat[1] {\delta\hat{C}_{#1}}
\newcommand \DChat[1] {\Delta\hat{C}_{#1}}

\newcommand \bcc {(\delta N)}

\begin{document}

\title{Density Functional Theory of Freezing and Phase Field Crystal Modeling}

\author{K. R. Elder$^1$, Nikolas Provatas$^2$, Joel Berry$^{1,3}$, Peter Stefanovic$^{2}$,
and Martin Grant$^{3}$ }

\affiliation{$^1$Department of Physics, Oakland University,
Rochester, MI, 48309-4487} \affiliation{$^2$ Department of Materials
Science and Engineering and Brockhouse Institute for Materials
Research, McMaster University, Hamilton, ON, Canada L8S-4L7}
\affiliation{$^3$ Physics Department, Rutherford Building, 3600 rue
University, McGill University, Montr\'eal, Qu\'ebec, Canada H3A 2T8}

\date{\today}

\begin{abstract}

In this paper the relationship between the density functional theory of
freezing and phase field modeling is examined.  More specifically
a connection is made between the correlation functions that enter
density functional theory and the free energy functionals used in
{\it phase field crystal} modeling and standard models of binary alloys
(i.e., regular solution model). To demonstrate the properties of the
phase field crystal formalism a simple model of binary alloy
crystallization is derived and shown to simultaneously model
solidification, phase segregation, grain growth, elastic and plastic
deformations in anisotropic systems with multiple crystal
orientations on diffusive time scales.

\end{abstract}

\pacs{05.70.Ln, 64.60.My, 64.60.Cn, 81.30.Hd}
\maketitle

\vskip1pc

\tableofcontents

\section{Introduction}

    The formalism for calculating equilibrium states was established
many years ago by Gibbs, Boltzmann and others. While this formalism
has proved remarkably successful there are many systems which never
reach equilibrium, mainly due to the existence of metastable or long
lived transient states. This is most apparent in solid materials.
For example it is very unlikely that the reader is sitting
a room containing any single crystals except items produced
with considerable effort such as the silicon chips in computers.
In fact the vast majority of naturally occurring or
engineered materials are not in equilibrium and contain complex
spatial structures on nanometer, micron or millimeter length scales.
More importantly many material properties (electrical, optical,
mechanical, etc.) are strongly influenced by the non-equilibrium
structures that form during material processing. For example the
yield strength of a polycrystal varies as the inverse square of the
average grain size.

    The study of non-equilibrium microstructure formation has seen
considerable advances through the use of the {\it phase field}
approach. This methodology models the dynamics of various continuum
fields that collectively characterize microstructure in phase
transformations. For example, phase field or continuum models have been
used to simulate spinodal decomposition \cite{CHC}, order-disorder
transition kinetics \cite{ac75}, ordering of block-copolymer
melts \cite{bcp}, solidification of pure and binary
systems \cite{Col85,Lan86,gegk93,edkg94,War95} and many other systems.
In these phenomena the evolution of the appropriate field(s) (e.g.,
solute concentration in spinodal decomposition) is assumed to be
dissipative and driven by minimizing a phenomenological free energy
functional \cite{CHC}.

    Advances in the phase field modeling of solidification
phenomena have followed a progression of innovations, beginning with
the development of free energies that capture the thermodynamics of
pure materials \cite{Col85,Lan86,gegk93} and alloys
\cite{edkg94,War95}. Several modification were then proposed
\cite{Kob93,Cag92,Wan93} to simplify numerical simulations and
improve computational efficiency.  Perhaps the most important
innovation was the development of matched asymptotic analysis
techniques that directly connect phase field model parameters with
the classical Stefan (or sharp-interface) models for pure materials
or alloys \cite{Kar96,Kar01,egpk01,Pro05}. These techniques were
complimented by new adaptive mesh refinement algorithms
\cite{Pro98,Pro99}, whose improved efficiency significantly
increased the length scales accessible by numerical simulations,
thus enabling the first experimentally relevant simulations of
complex dendritic structures and their interactions in organic and
metallic alloys \cite{Pro99b,Pro03,Lan03,Gre04,Lan04}.

    A weakness of the traditional phase field methodology is
that it is usually formulated in terms of fields that are spatially
uniform in equilibrium.  This eliminates many physical features that
arise due to the periodic nature of crystalline phases, including
elastic and plastic deformation, anisotropy and multiple
orientations.   To circumvent this problem traditional phase field
models have been augmented by the addition of one or more auxiliary
fields used to describe the density of dislocations \cite{wjck01,jk01,hmrg02}
continuum stress and strain fields \cite{ck91,wk97} and orientation
fields \cite{cy94,mesg95,Kob00}. These approaches have proven quite
useful in various applications such as
polycrystalline solidification \cite{jk01,cy94,mesg95,Kob00,War03,Gra04}.
Nevertheless it has proven quite challenging to incorporate elasto-plasticity,
diffusive phase transformation kinetics and anisotropic surface
energy effects into a single, thermodynamically consistent model.

    Very recently a new extension to phase field modeling has
emerged known as the {\it phase field crystal} method (PFC)
\cite{ekhg02,eg04,bme06}.  This methodology describes the evolution
of the atomic density of a system according to dissipative dynamics
driven by free energy minimization.  In the PFC approach the free
energy functional of a solid phase is minimized when the density
field is periodic. As discussed in prior publications
\cite{ekhg02,eg04,bme06} the periodic nature of the density field
naturally gives rise to elastic effects, multiple crystal
orientations and the nucleation and motion of dislocations. While
these physical features are included in other atomistic approaches
(such as molecular dynamics) a significant advantage of the PFC
method is that, by construction, it is restricted to operate on
diffusive time scales not on the prohibitively small time scales
associated with atomic lattice vibrations.   A similar approach 
has also been recently proposed by Jin and Khachaturyan \cite{jk06}.
In the case of pure materials the PFC approach has been shown \cite{ekhg02,eg04} to
model many phenomena dominated by atomic scale elastic and plastic
deformation effects. These include grain boundary interactions,
epitaxial growth and the yield strength of nano-crystals.  

    The original PFC model is among the simplest mathematical
descriptions that can self-consistently combine the physics of
atomic-scale elasto-plasticity with the diffusive dynamics of phase
transformations and microstructure formation.   Nevertheless,
analogously to traditional phase field modeling of solidification,
further work is required to fully exploit the methodology.  More
specifically it is important to be able to generalize the method to
more complex situations (binary alloys, faster dynamics, different
crystal structures, etc.), to develop more efficient numerical
techniques and to make a direct connection of the parameters of the
model to experimental systems.  Several innovations toward this goal
have already been developed. Goldenfeld {\it et al.}
\cite{agd05a,agd05b} have recently derived amplitude equations for
the PFC model which are amenable to adaptive mesh refinement
schemes. This work has the potential to enable simulations of
mesoscopic phenomena ($\mu m \rightarrow mm$) that are resolved down
to the atomic scale and incorporate all the physics discussed above.
Another recent advance is the inclusion of higher order time
derivatives in the dynamics to simulate ``instantaneous'' elastic
relaxation \cite{Ste06}. This extension is important for modeling
complex stress propagation and externally imposed strains. Very
recently, Wu {\it et al.} \cite{wk06} fitted the PFC parameters to
experimental data in iron and were able to to show that the PFC
model gives an accurate description of the anisotropy of the surface
tension. In addition to this work Wu and Karma have also developed a
simple and elegant scheme to extend the method to other crystal
symmetries (i.e., FCC in three dimensions).

    The purpose of this paper is to link the formalism of density
function theory (DFT) of freezing, as formulated by Ramakrishnan and
Yussouff \cite{dft0} (and also reviewed by many other authors, such
as Singh \cite{dft1}) with the phase field crystal (PFC) method and
to exploit this connection to develop a phase field crystal
model for binary alloys. The organization of the paper and a summary
of the remaining sections is as follows.

    In Section IIA the density functional theory of freezing of
pure and binary systems is briefly outlined.  In this approach the
free energy functional is written in terms of the time averaged
atomic density field $\rho$ ($\rho_A$ and $\rho_B$ in binary
systems) and expanded around a liquid reference state existing along
the liquid/solid coexistence line. Formally the expansion contains
the n-point correlation functions of the liquid state.  In this work
the series expansion of the free energy is truncated at the 2-point
correlation function, $C(\vec{r}_1,\vec{r}_2)$.

Within this framework it is shown in Section IIIA that the PFC model
for a pure material can be recovered from DFT if
$C(\vec{r}_1,\vec{r}_2)$
 is parameterized by three constants
related to the liquid and solid state compressibilities and the
lattice constant. The parameters of the PFC model can thus be directly
related to the physical constants that enter the DFT of freezing and
the PFC model can be viewed as a simplified form of DFT.  In Section
IIIB a binary system is considered.  Similar to the case of
pure materials the free energy expansion of a binary alloy will be
truncated at the 2-point correlation functions which are then
characterized by three parameters.  At this level of simplification
it is shown that the ``regular" solution model used in materials physics
for alloys
can be obtained directly from density functional
theory. It is shown that the phenomenological nearest
neighbour bond energies that enter the ``regular" solution model are
equal to the compressibilities that enter density functional theory.
This section also provides insight into the concentration dependence
of various properties of the crystalline phase of a binary alloy
such as the lattice constant, effective mobilities and elastic
constants.

In Section IIIC a simplified version of the binary
alloy free energy is derived. This is done in order to provide a
mathematically simpler model that can more transparently illustrate
the use of the PFC formalism in simultaneously modeling diverse
processes such as solidification, grain growth, defect nucleation,
phase segregation and elastic and plastic deformation. This section
also shows that the free energy of the simplified alloy PFC model
reproduces two common phase diagrams associated with typical binary
alloys in materials science.

    In Section IV dynamical equations of motion that govern the
evolution of the solute concentration and density field of the
binary alloy are derived. Finally in Section V the simplified
binary alloy model is used to simulate several important
applications involving the interplay of phase transformation
kinetics and elastic and plastic effects. This includes
solidification, epitaxial growth and spinodal decomposition. Some of
the more tedious calculations are relegated to the Appendices.

\section{Density Functional Theory of Freezing}

    In this section free energy functionals of pure and
binary systems as derived from the density functional theory of freezing are
presented. A derivation of the functional for a pure
materials is outlined in the Appendix \ref{app:DFT}. For a more
rigorous treatment the reader is referred to the
work of Ramakrishnan and Yussouff \cite{dft0} and numerous other
very closely related review articles by Singh \cite{dft1}, Evans
\cite{Eva79} and references therein.

\subsection{Single Component System}
\label{DFT:single}

    In density functional theory the emergence of an ordered phase
during solidification can be viewed as a transition to a phase
in which the atomic number density, $\rho(\vec{r})$, is
highly non-homogenous and possesses the spatial
symmetries of the crystal \cite{dft1}. This approach implicitly
integrates out phonon modes in favour of a statistical
view of the ordered phase that changes on diffusive
time scales. The free energy functional of a system is expressed
in terms of $\rho$ and constitutes the starting point of the PFC
model.

\begin{figure}[btp]
\center{\includegraphics*[width=0.53\textwidth]{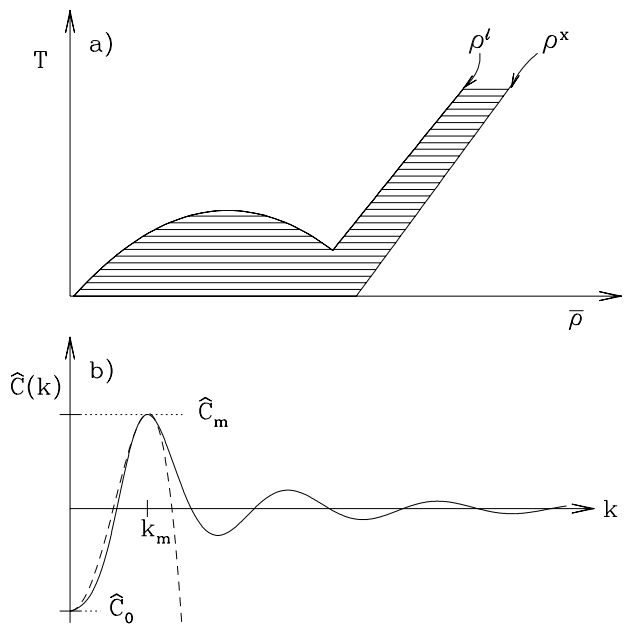}} \vskip
-1in \caption{(a) Sample phase diagram. In this figure the shaded
area corresponds to a coexistence region.  In the calculations
presented in this paper the correlation functions are taken from
points along the liquidus line at density $\rho^{\ell}$.  (b) In
this figure a `typical' liquid state two-point direct correlation
function is sketched.  The dashed line represents the approximation
used in most of the manuscript.} \label{fig:phdiC2n}
\end{figure}

    In this work the free energy functional, denoted ${\cal F}[\rho]$,
is expanded functionally about a density, $\rho=\rho_l$,
corresponding to a liquid state lying on the liquidus line of the
solid-liquid coexistence phase diagram of a pure material as shown
in Fig. (\ref{fig:phdiC2n}a). The expansion is performed in powers
of $\delta \rho \equiv \rho-\rho_l$.

As shown by other others and outlined in Appendix \ref{app:DFT}
the free energy density can be written as
\ben
\frac{{\cal F}_c}{k_B T}&=&\int d\vec{x} \left[ \rho(\vec{r}) \ln
\left( \frac{\rho(\vec{r})}{\rho_l}\right)
-\delta \rho(\vec{r}) \right] \nonumber \\
-\frac{1}{2} &\int& d\vec{r}_1 d\vec{r}_2 \delta \rho(\vec{r}_1)
C_{2}(\vec{r}_1, \vec{r}_2) \delta \rho(\vec{r}_2) + \nonumber \\
-\frac{1}{6} &\int& d\vec{r}_1 d\vec{r}_2 d\vec{r}_3 \delta
\rho(\vec{r}_1) C_{3}(\vec{r}_1, \vec{r}_2, \vec{r}_3) \delta
\rho(\vec{r}_2) \delta \rho(\vec{r}_3) \nonumber \\ + &\cdots &
\label{free_eng2}
\een
where ${\cal F}_c$ is the free energy corresponding to the density
$\rho(\vec{r})$ minus that at the constant density $\rho_l$. The
function, $C_{2}$ is the two point direct correlation function of an
isotropic fluid. It is usually denoted $C_{ij} \equiv C_{2}(r_{12})$
and satisfies $r_{12}\equiv |\vec{r}_1-\vec{r}_2|$. The function
$C_{3}$ is the three point correlation function, etc.. Formally the
correlation functions are defined by
\ben
C_1(\vec{r}) &\equiv & \frac{\delta \Phi[\rho(\vec{r})]}{\delta
\rho(\vec{r})} \nline C_2(\vec{r}_1,\vec{r}_2) &\equiv&
\frac{\delta^2 \Phi}{\delta \rho(\vec{r}_1) \delta \rho(\vec{r}_2)}
\nline C_3(\vec{r}_1,\vec{r}_2,\vec{r}_3) &\equiv& \frac{\delta^3
\Phi}{\delta \rho(\vec{r}_3) \delta \rho(\vec{r}_1) \delta
\rho(\vec{r}_2)}  \nline && \cdots \label{C_funcs}
\een
where $\Phi[\rho]$ represents the total potential energy of
interactions between the particles in the material. The proof that
$\Phi$ is a functional of $\rho$ is shown rigorously in Evans
\cite{Eva79}.

\subsection{Multi-Component System}

    For an alloy involving one or more components the free energy
functional of a pure material in Eq. (\ref{free_eng2}) is extended
to the form
\ben
\frac{{\cal F}_c}{k_B T} &=& \sum_{i} \int d\vec{r}\,
\Big[\rho_i\left(\vec{r}\right) \ln{\left(\rhoi
\left(\vec{r}\right)/\liqi\right)} - \drhoi\left(\vec{r}\right)\Big]
\nline &-& \frac{1}{2}\sum_{i,j} \int d\vec{r}_1\,d\vec{r}_2\,
\drhoi(\vec{r}_1)\,C_{ij}\,(\vec{r}_1,\vec{r}_2)\, \drhoj(\vec{r}_2)
+ \cdots \nline \label{multi_F}
\een 
where the sums are over the elements in the alloy, $\drhoi \equiv
\rhoi-\liqi$, $\liqi$ is the value of the number density of
component $i$ on the liquid side of the liquid/solid coexistence
line. The function $C_{ij}$ is the two point direct correlation
function of between components $i$ and $j$ in an isotropic fluid. As
in the case of a pure materials it will be assumed that $C_{ij}
\equiv C_{ij}(r_{12})$, where $r_{12}\equiv |\vec{r}_1-\vec{r}_2|$.
The next term in the expansion of Eq.~(\ref{multi_F}) contains the
three point correlation, the next after that, the four point, etc..
In this paper only two point correlations will be considered, but it
must be stressed that these higher order correlations maybe crucial
for some systems, such as Si.   Before considering the properties of
the binary alloy free energy in detail it is instructive to first
study the properties of a pure system and show the connection
between this formalism and the original phase field crystal model.

\section{Analysis of Free Energy Functionals}

\subsection{Pure Materials}
\label{sec:pure}

    In this section the free energy functional of a single
component alloy is considered in the limit that the series
given in Eq. (\ref{free_eng2}) can be truncated at $C_2$, i.e.,
\ben
 {\cal F}/{k_B T} &=&
\int d\vec{r}\, \left[ \rho \ln \left({\rho}/{\liqr}\right) - \drho
\right] \nline &-& ({1}/{2}) \int d\vec{r}\,d\vec{r}'\, \drho\,
C(\vec{r},\vec{r}')\, \drho '\,
\een
where, for convenience, the subscript `$2$' has been dropped from
the two-point correlation function as has the subscript $c$ from
${\cal F}_c$.  To understand the basic features of this free energy
functional it is useful to expand ${\cal F}$ in the dimensionless
deviation of the density $\rho$ from its average, $\bar{\rho}$,
using the re-scaled density
\be 
n \equiv (\rho -\bar{\rho})/\bar{\rho}.
\ee 
Expanding ${\cal F}$ in powers of $n$ gives,
\ben
\frac{\Delta \cal F}{\bar{\rho}\,k_B T} &=& \int d\vec{r}\,
\left[n\frac{1-\bar{\rho}\,C}{2} n -\frac{n^3}{6}
+\frac{n^4}{12}-\cdots\right], \label{pure_free}
\een 
where $\Delta {\cal F} \equiv {\cal F} - {\cal F}_o$ and ${\cal
F}_o$ is the the free energy functional at constant density (i.e.,
$\rho =\bar{\rho}$). For simplicity $C$ is an operator defined such
that $n\, C\, n \equiv \int d\vec{r}\,' n(\vec{r})
C(|\vec{r}-\vec{r}\,'|) n(\vec{r}\,')$. Terms that are
linearly proportional to $n$ in the above integral are identically
zero by definition.

    To gain insight into the properties of the free energy
functional in Eq.~(\ref{pure_free}) it is useful to expand the two
point correlation function in a fourier series, i.e.,
\be
\hat{C} = \Chat{0} + \Chat{2} k^2 + \Chat{4} k^4 + \cdots.
\ee
(in real space this corresponds to $C=(\hat{C}_0-\hat{C}_2
\nabla^2+\hat{C}_4\nabla^4-\cdots)\delta
(\vec{r}-\vec{r}^{\prime})$, where the gradients are with respect to
$\vec{r}^{\prime}$). The function $\hat{C}$ is sketched for a
typical liquid in Fig. (\ref{fig:phdiC2n}b).  In what follows only
terms up to $k^4$ will be retained.  In this manner the properties
of the material are parameterized by the three variable, $\Chat{0}$,
$\Chat{2}$ and $\Chat{4}$.  To fit the first peak in $\hat{C}$,
$\Chat{0}$, $\Chat{2}$ and $\Chat{4}$ must be negative, positive and
negative, respectively.  These variables are related to three basic
properties of the material, the liquid phase isothermal
compressibility ($\sim (1-\bar{\rho}\Chat{0}$)), the bulk modulus of
the crystal ($\sim \bar{\rho}\,\Chat{2}^2/|\Chat{4}|)$ and lattice
constant ($\sim (\Chat{2}/|\Chat{4}|)^{1/2}$).  In other words the
$k=0$ term is related to the liquid phase isothermal
compressibility, the height of the first peak ($\hat{C}_m$ in Fig.
(\ref{fig:phdiC2n}b)) is related to the bulk modulus of the
crystalline phase and the position of the first peak determines the
lattice constant.

    It is important to note that at this level of simplification the
material is only defined by three quantities which may not be enough
to fully parameterize any given material.  For example this simple three
parameter model always predicts triangular symmetry in two
dimensions and BCC symmetry in three dimensions.   Other crystal
symmetries can be obtained by using more complicated 2-point
correlation functions \cite{wk06} or by including higher order
correlation functions.

    In two dimensions $\cal F$ is minimized by a triangular lattice that
can be represented to lowest order by a one-mode approximation as
\be 
n = A \left(
\frac{1}{2}\cos \left(\frac{2qy}{\sqrt{3}}\right)- \cos (qx) \, \cos
\left(\frac{qy}{\sqrt(3)}\right) \right).
\label{eq:dimden}
\ee 
Substituting Eq. (\ref{eq:dimden}) into Eq. (\ref{pure_free}) gives,
\ben
\Delta F &=& \frac{3}{16}(1-\bar{\rho}\,\Chat{0})\,A^2 -
\frac{1}{32}A^3 + \frac{15}{512}A^4 - \cdots \nline &&
-\bar{\rho}\,\left( \frac{1}{4} \Chat{2}\, q^2 +\frac{1}{3}
\Chat{4}\, q^4 + \cdots \right)\,A^2
\een
where $\Delta F \equiv \Delta {\cal F}/(\bar{\rho}\,k_B\,T\,S)$ and $S$
is the area of a unit cell.
Minimizing $\Delta F$ with respect to $q$ gives the equilibrium
wavevector, $q_{eq}$, which is
\be
q_{eq} = \sqrt{ 3 \Chat{2} /(8 |\Chat{4}|) }. \label{eq:qstar}
\ee 
or in terms of the equilibrium lattice constant $a_{eq} =
2\pi/q_{eq}$.
When $q=q_{eq}$, $\Delta \cal F$ becomes
\ben 
\Delta F &=& \frac{3}{16}\,\Delta B\,A^2-\frac{1}{32}\,A^3
+\frac{15}{512}\,A^4 +\cdots, \label{eq:dfpure}
\een 
where $\Delta B \equiv B^{\ell}-B^s$, $B^{\ell} \equiv 1-\bar{\rho}
\Chat{0}$ and $B^s \equiv \bar{\rho}
\,(\Chat{2})^2/({4|\Chat{4}|})$.  The parameter
$B^{\ell}$ is the dimensionless bulk modulus of the liquid state
(i.e., $B^{\ell} = \kappa/(\bar{\rho}\,k_B\,T)$, where $\kappa$ is the
bulk modulus of a liquid).  The parameter $B_s$ is
proportional to the bulk modulus in the crystalline phase.

    Equation (\ref{eq:dfpure}) indicates that the liquid state is
linearly unstable to the formation of the crystalline phase when
$\Delta B < 0$.  This instability arises from a competition between
the elastic energy stored in the liquid and crystalline phases. It
is interesting to note that $\Delta B$ can also be written,
\be
\Delta B = (\bar{\rho}_s - \bar{\rho})/\bar{\rho}_s
\ee
where $\bar{\rho}_s = 1/\Chatm$ and $\Chatm$ is the height of the
first peak of $\hat{C}$ as shown in Fig. (\ref{fig:phdiC2n}).
Written in this form $\bar{\rho}_s$ can be thought of as defining
the effective spinodal density, i.e., the average density at which the
liquid becomes linearly unstable to crystallization.

    Unfortunately it is difficult to obtain the equilibrium state (i.e.,
by solving $d\Delta F /dA=0$) without truncating the infinite series
in Eq. (\ref{eq:dfpure}).  If only terms to order $A^4$ are retained
an analytic approximation can be obtained for the amplitude
($A_{min}$) that minimizes $\cal F$. In this approximation the
solution is
\ben
A_{min} = 2\left(1+ \sqrt{20\bar{\rho}/\bar{\rho}_s-19}\,\right)/5.
\label{eq:Amin1}
\een
Thus solutions for a crystalline state exist when
$\bar{\rho}\,>\,19/20\,\bar{\rho}_s$.

    It is also straightforward to calculate the change in energy of the
crystalline state upon deformation (i.e., bulk, shear or
deviatoric). Details of similar calculations are given in previous
publications \cite{ekhg02,eg04}.  The result of these
two-dimensional calculations gives the dimensionless bulk modulus,
$B_c$, of the crystalline phase, i.e.,
\be
B_c = \frac{3}{32} \bar{\rho} \frac{(\Chat{2})^2}{|\Chat{4}|}
A_{min}^2 = \frac{3}{8}\,B^s\,A_{min}^2. \label{eq:bcrys}
\ee
In other words the parameter $B^s$ controls the bulk modulus of the
crystalline phase.

    These calculations can be easily extended to three
dimensions.  As discussed previously this particular approximation
for $\hat{C}$ leads to a BCC crystal in three dimensions which can
be represented in a one mode approximation as,
\ben 
n &=& A\big(\cos{(qx)}\cos{(qy)}
 +\cos{(qx)}\cos{(qz)} \nline &&
+\cos{(qy)}\cos{(qz)}\big).
\een 
Substituting this functional form into the free energy and minimizing
with respect to $q$ gives,
\be
q_{eq}^{3d} = \sqrt{\hat{C}_2/|\hat{C}_4|},
\ee
and the free energy functional at this $q$ is,
\be
\Delta F^{3d} = \frac{3}{8}\Delta B A^2 - \frac{1}{8}A^3
+ \frac{45}{256}A^4 + \cdots .
\ee

    Thus in this instance the 'spinodal' occurs at the same density
as in the two dimensional case. If the series is truncated at $A^4$
the amplitude that minimizes the free energy is then
\be 
A_{min}^{3d} = 4\left(1+\sqrt{15\bar{\rho}/\bar{\rho}_s-14}\right)/15.
\ee 
Thus in this approximations crystalline (BCC) solutions only exist
if $\bar{\rho}\,>\,14\bar{\rho}_s$.  In addition the elastic
constants can also be calculated in 3D in the usual manner. For
example the dimensionless bulk modulus of the crystalline state is
given by; \be B_c^{3d} = 3\, B^s (A_{min}^{3d})^2. \ee This
calculation gives the basic functional dependence of the
(dimensionless) bulk modulus on $B^s$ and the amplitude.  For a more
accurate calculation higher order fourier components and more terms
in the powers series in $n$ should be retained.

    Finally it is useful to consider fixing the density and varying the
temperature. If the liquidus and solidus lines are roughly linear
then, $\bar{\rho}_s$ can be approximated by a linear function of
temperature.  In the sample phase diagram shown in Fig.
(\ref{fig:phdiC2n}a) the liquidus and solidus lines are roughly
parallel and it is likely that the spinodal is also roughly parallel
to these lines.  In this case $\Delta B$ can be written, \be
\label{eq:DBT} \Delta B = \alpha \Delta T \ee where $\Delta T \equiv
(T-T_s)/T_s$, $T_s$ is the spinodal temperature and $\alpha \equiv
(T_s/\bar{\rho}_s)(\partial \bar{\rho}_s/\partial T_s)$, evaluated
at $\rho=\bar{\rho}$.

\subsection{Binary Alloys}
\label{sect:alloys}

    For a binary alloy made up of `$A$' and `$B$' atoms the free energy
functional can be written to lowest order in terms of the direct
correlation functions as,
\ben 
\frac{{\cal F}}{k_BT} &=&\int
d\vec{r} \left[ \rho_A \ln\left(\frac{\rhoA}{\liqA}\right) -\drhoA +
\rho_B \ln\left(\frac{\rhoB}{\liqB}\right) -\drhoB\right] \nline &&
-\frac{1}{2}\int d\vec{r}_1 d\vec{r}_2\Big[\drhoA(\vec{r}_1)\,
C_{AA}(\vec{r}_1,\vec{r}_2)\, \drhoA(\vec{r}_2) \nline && \ \ \ \ \
\ \ \ \ \ \ \ \ \ +\drhoB(\vec{r}_1)\, C_{BB}(\vec{r}_1,\vec{r}_2)\,
\drhoB(\vec{r}_2) \nline &&
 \ \ \ \ \ \ \ \ \ \ \ \ \ \
+2\,\drhoA(\vec{r}_1)\, C_{AB}(\vec{r}_1,\vec{r}_2)\,
\drhoB(\vec{r}_2) \Big] \label{full_free_eng}
\een 
where $\delta \rho_A \equiv\rho_A-\rho_l^A$ and $\delta \rho_B
\equiv \rho_B-\rho_l^B$. It is assumed here that all two point
correlation functions are isotropic, i.e.,
$C_{ij}(\vec{r}_1,\vec{r}_2) = C_{ij}(|\vec{r}_1-\vec{r}_2|)$.

    In order to make a connection between the alloy free energy and
standard phase field models it is useful to define the
total number density, $\rho \equiv
\rhoA+\rhoB$ and a local concentration field $c \equiv \rhoA/\rho$.
In terms of these fields the atomic densities can be written,
$\rho_A=c \rho$ and $\rho_B= \rho(1-c)$. Furthermore it is useful to
define $\rho=\rho_l+\delta \rho$ where $\rho_l \equiv
\rho_l^A+\rho_l^B$ and $\delta c=1/2-c$. Substituting these
definitions into Eq. (\ref{full_free_eng}) gives,
\ben 
\frac{{\cal F}}{\,k_BT} &=&\int d\vec{r} \Big[  \rho\ln (\rho/\liqr)
- \drho + \nline &&-\frac{1}{2}\drho\,\big\{c\, C_{AA}+(1-c)\,
C_{BB}\big\}\,\drho \nline && + \rho \big\{c\ln\left(c\right)
+\left(1-c\right)\ln\left(1-c\right) \big\} \nline && + \rho c
\Delta C \, (1-c) \rho +\beta \, \dc +F_o \Big] \label{eq:freeb}
\een 
where
\ben 
\Delta C &\equiv& (C_{AA}+C_{BB})/2-C_{AB}, \nline \beta &\equiv&
\frac{\rho_l}{2} (C_{AA}-C_{BB}) (\rho+\rho_l) + \rho\,\ln \left(
\frac{\liqB}{\liqA} \right)
\een
and
\ben
\nline F_o &\equiv & \rho
\ln\left(\frac{\liqr}{\sqrt{\liqA\liqB}} \right) - \frac{C_{AA}}{2}
\left((\liqA)^2+\frac{\rho_l}{2}(\rho_l+\rho) \right) \nline && -
\frac{C_{BB}}{2}
\left((\liqB)^2+\frac{\rho_l}{2}(\rho_l+\rho)\right).
\label{ally_free_eng1}
\een 

    To illustrate the properties of the model in Eq.~(\ref{eq:freeb})
it useful to consider two limiting cases, a liquid phase at constant
density and a crystalline phase at constant concentration. These
calculations are presented in following two sub-sections.

\subsubsection{Liquid phase properties}

    In the liquid phase $\rho$ is constant on average and in the
mean field limit can be replaced by $\rho = \bar{\rho}$.
To simplicity calculations, the case $\rho = \bar{\rho} \approx \liqr$
(or $\drho \approx 0$) will now be considered.  As in
the previous section it is
useful to expand the direct correlation functions in Fourier space, i.e.,
\ben 
\hat{C}_{ij} = \cij{0} +\cij{2}k^2+\cij{4}k^4 + \cdots  .
\een 
where the subscript $i$ and $j$ refer to a particular element.
Substituting the real-space counterpart of the Fourier expansion for
$\hat{C}_{ij}$ (to order $k^2$) into Eq. (\ref{eq:freeb}) gives,
\ben 
\frac{{\cal F}_C}{\bar{\rho}\,k_BT}&=&\int
d\vec{r}\,\Big[c\ln\left(c\right)
+\left(1-c\right)\ln\left(1-c\right) \nline &+&
\frac{\bar{\rho}\DChat{0}}{2}\,c(1-c) +\gamma^{\ell}\,\delta c +
\frac{\bar{\rho}\DChat{2}}{2}|\nabla c|^2 \Big], \label{eq:regular}
\een 
where ${\cal F}_C$ is the total free energy minus a constant that
that depends only on $\bar{\rho}$, $\rho_A^l$ and $\rho_B^l$,
\be 
\gamma^{\ell} \equiv (B^{BB}_{\ell}-B^{AA}_{\ell})+ \bar{\rho}\,\ln
\left(\liqB/\liqA\right),
\label{eq:gam}
\ee 
\be
\DChat{n}\equiv \cAA{n}+\cBB{n}-2\cAB{n}.
\ee 
and $B^{ij}_{\ell}=1-\bar{\rho}\hat{C}^{ij}_0$ is the dimensionless
bulk compressibility. Equation (\ref{eq:regular}) is the regular solution
model of a binary alloy.

    The coefficient of $c(1-c)$ in Eq.~(\ref{eq:regular}) is given by
\be 
\bar{\rho} \Delta \hat{C}_0 = 2
B^{AB}_{\ell}-B^{AA}_{\ell}-B^{BB}_{\ell}.
\ee 
This result shows that in the liquid state the 'interaction'
energies that enter regular solution free energies are simply the
compressibilities (or the elastic energy) associated with the atomic
species.   The $\gamma^{\ell}$ term is also quite interesting as it
is responsible for asymmetries in the phase diagram.
Thus Eq. (\ref{eq:gam}) implies that asymmetries can arise from
either different compressibilities or different densities.

    Expanding Eq. (\ref{eq:regular}) around $c=1/2$ gives,
\ben 
\frac{\Delta {\cal F}_C}{\bar{\rho}\,k_B T} =\int
d\vec{r}\,\Big[\frac{r^{\ell}}{2} \dc^2 + \frac{u}{4} \dc^4 +
\gamma^{\ell}\,\delta c + \frac{K}{2}|\nabla c|^2 \Big]
\een 
where, $\Delta {\cal F}_C \equiv {\cal F}_C - \bar{\rho} k_B T \int
d\vec{r}(\bar{\rho}\DChat{0}/8-\ln (2) )$, $u \equiv 16/3$,
$r^{\ell} \equiv (4 - \bar{\rho} \DChat{0})$ and $K=\bar{\rho}
\DChat{2}$. The parameter $r^{\ell}$ is related only to the $k=0$
part of the two-point correlation function and can be written,
\be 
r^{\ell} = 4 +(B_{\ell}^{AA}+B_{\ell}^{BB}-2B_{\ell}^{AB}).
\ee 
This result implies that the instability to phase segregation in the fluid
is a competition between entropy ($4$) and the elastic energy of a
mixed fluid ($2B_{\ell}^{AB}$) with the elastic energy associated
with a phase separated fluid ($B_{\ell}^{AA}+B_{\ell}^{BB}$).
Replacing the dimensionless bulk moduli with the dimensional version
(i.e., $B = \kappa/k_B T$), gives the critical point (i.e. $r^l=0$)
as
\be
 T^{\ell}_C = (2 \kappa_{AB}-\kappa_{BB}-\kappa_{AA})/(4k_B).
\ee
 where $\kappa$ is the dimensional bulk modulus.

    The properties of the crystalline phase are more complicated but at
the simplest level the only real difference is that the elastic
energy associated with the crystalline state must be incorporated.
This is discussed in the next section.

\subsubsection{Crystalline phase properties}

    To illustrate the properties of the crystalline state, the case in
which the concentration field is a constant is considered.  In this
limit the free energy functional given in Eq. (\ref{eq:freeb}) can
be written in the form
\ben 
\frac{{\cal F}}{k_B T}  \equiv \int d\vec{r} \left[\rho \ln
\left(\frac{\rho}{\liqr}\right) -\dr -\frac{1}{2}\dr\,\bar{C}\,\dr +
G\right]
\een 
where $G$ is a function of the concentration $c$ and $\rho_l$ and
couples only linearly to $\drho$. The operator $\bar{C}$ can be
written as
\ben 
\bar{C} \equiv  c^2 C_{AA} + (1-c)^2 C_{BB} + 2c(1-c) C_{AB}.
\label{eq:cbar}
\een 
Thus in the limit that the concentration is constant this free
energy functional is that of a pure material with an effective two
point correlation function that is an average over the $AA$, $BB$
and $AB$ interactions.  In this limit the calculations presented in
section \ref{sec:pure} can be repeated using the same approximations
(i.e., expanding $\rho$ around $\liqr$, expanding $\bar{C}$ to
$\nabla^4$ and using a one mode approximation for $\dr$) to obtain
predictions for the concentration dependence of various quantities.
For example the concentration dependence of the equilibrium
wavevector (or lattice constant, Eq. (\ref{eq:qstar})) and bulk
modulus Eq. (\ref{eq:bcrys}) can be obtained by substituting
$\Chat{n} = c^2\cAA{n}+(1-c)^2\cBB{n}+2c(1-c)\cAB{n}$.

    As a more specific example the equilibrium lattice constant can be
expanded around $c=1/2$ to obtain in two or three dimensions, \be a_{eq}(\dc) =
a_{eq}(0)\left(1+\eta\, \dc + \cdots\right) \ee where $\dc = c-1/2$
and $\eta$ is the solute expansion coefficient given by,
\be
\eta = (\dChat{4}-\dChat{2})/2 \ee where \be \dChat{n} \equiv
(\cAA{n}-\cBB{n})/\Chat{n}^o
\ee 
and $\Chat{n}^o \equiv
\Chat{n}(\dc=0)=(\cAA{n}+\cBB{n}+2\cAB{n})/4$.

    This line of reasoning can also be used to understand the influence
of alloy concentration on crystallization. Specifically, for the
case of an alloy, the terms in Eq. (\ref{eq:dfpure}) (with $A$
replaced with $A_{min}$) become functions of concentration, since
$\Delta B$ and $A_{min}$ are concentration dependent.  Here, $\Delta
B$ can be expanded around $c=1/2$, i.e.,
\be
\Delta B(\dc) = \Delta B_0+\Delta B_1\, \dc + \Delta B_2\,\dc^2 +
\cdots
\ee
where $\Delta B_0=B_0^l-B_0^s$, $\Delta B_1=B_1^l-B_1^s$ and $\Delta
B_2=B_2^l-B_2^s$ are determined in the appendix. This would imply
that in the crystalline phase the free energy has a term of the
form, $r^c (\delta c)^2$, where
\ben
r^c &=& r^{\ell} + \frac{3\Delta B_2}{8} A_{min}^2 \nline &=&
4-\bar{\rho} \DChat{0}(1+ 3 A_{min}^2/8) -3/8 B^s_{2}A_{min}^2,
\een
in two dimensions (in three dimensions the $3/8$ factor is
replaced with $3/4$).
This result indicates that crystallization (i.e., a non-zero $A_{min}$)
favours phase segregation, assuming $\kappa_{AA}+\kappa_{BB} <
2\kappa_{AB}$.  For example, when $B^s_2=0$, the critical temperature increases
and can be written,
\be
T^c_C = T^{\ell}_C (1 + 3A_{min}^2/8),
\ee
or $T^c_C = T^{\ell}_C (1 + 3A_{min}^2/4)$ in three dimensions.

\subsection{Simple Binary Alloy Model}
\label{sec:simple}

    In this section a simple binary alloy model is
proposed based on a simplification of the free energy in
Eq.~(\ref{eq:freeb}). The goal of this section is to develop a
mathematically {\it simple} model that can be used to simultaneously
model grain growth, solidification, phase segregation in the
presence of elastic and plastic deformation. To simplify
calculations it is convenient to first introduce the following
dimensionless fields,
\ben 
n_A&\equiv&(\rhoA-\bar{\rho}_A)/\bar{\rho} \nline
n_B&\equiv&(\rhoB-\bar{\rho}_B)/\bar{\rho}.
\label{simple:field_defs1}
\een 
Also, it is convenient to expand in the following two fields,
\ben
n&=& n_A+n_B \nline \delta N
&=&(n_B-n_A)+\frac{\bar{\rho}_B-\bar{\rho}_A}{\bar{\rho}} .
\label{simple:field_defs2}
\een 
The following calculations will use the field $\delta N$ instead of
$\delta c$.  Expanding Eq. (\ref{eq:freeb}) around $\delta N=0$ and
$n=0$ gives a free energy of the form
\ben
\label{eq:free}
\frac{{\cal F}}{\bar{\rho}\, k_B T} &=& \int
d\vec{r} \Big(\frac{n}{2}\left[B_{\ell}+ B_s\left( 2R^2\nabla^2
+R^4\nabla^4\right)\right]n \nline && +\frac{t}{3}n^3
+\frac{v}{4}n^4 + \gamma\dN+\frac{w}{2} \dN^2 + \frac{u}{4} \dN^4
\nline && + \frac{L^2}{2} |\vec{\nabla}\dN|^2 + \cdots \Big).
\een 
Details of this free energy and explicit expressions for $B^\ell$,
$B^s$ and $R$ are given in Appendix \ref{app:simple}.  For
simplicity the calculations presented in this section are for
a two dimensional system.

The transition from liquid to solid is intimately
related to $\Delta B=B^{\ell}-B^s$ as was the case for the pure
material and can be written in terms of a temperature difference,
i.e., Eq. (\ref{eq:DBT}).   In addition some of the polynomial
terms in $n$ and $\delta N$ have been multiplied by variable
coefficients even though they can be derived exactly as shown in Appendix
\ref{app:simple}. For example the parameter $v=1/3$ recovers the
exact form of the $n^4$ term. This flexibility in the choice of
coefficients was done to be able to match the parameter of the free
energy with experimental materials parameters. As an example Wu and
Karma \cite{wk06} showed that adjusting the parameter $v$ can be
used to match the amplitude of fluctuations obtained in molecular
dynamics simulations.  With this fit they are able to accurately
predict the anisotropy of the surface energy of a liquid/crystal interface
in iron.

    To facilitate the calculation of the lowest order phase diagram
corresponding to Eq. (\ref{eq:free}) it is convenient to assume the
concentration field $\delta N$ varies significantly over length
scales much larger than the atomic number density field $n$. As a
result, the density field can be integrated out of the free energy
functional.  In this instance the one-mode approximation for $n$,
i.e.,
\be 
n = A \left( \frac{1}{2}\cos
\left(\frac{2qy}{\sqrt{3}}\right)- \cos (qx) \, \cos
\left(\frac{qy}{\sqrt(3)}\right) \right).
\label{eq:onag}
\ee 
will be used.
Substituting Eq. (\ref{eq:onag}) into  Eq. (\ref{eq:free}) and minimizing with
respect to $q$ and $A$ (recalling that $\delta N$ is assumed
constant over the scale that $n$ varies) gives
\be 
q_{eq} = \sqrt{3}/(2R) \ee and \be A_{min} =
\frac{4}{15v}\left(t+\sqrt{t^2-15v\Delta B}\right).
\ee 
The free energy that is minimized with respect to amplitude and
lattice constant is then,
\ben 
F_{sol}&=& \frac{w}{2}\dN^2 + \frac{u}{4}\dN^4 +\frac{3}{16}\Delta B
A_{min}^2 -\frac{t}{16} A_{min}^3 \nline && +\frac{45v}{512}
A_{min}^4.
\een 

    For mathematical simplicity all further calculations will
be limited to the approximations
$B^{\ell}=B^{\ell}_0+B^{\ell}_2(\dN)^2$ and $B^s=B^s_0$.  In this
limit analytic expression can be obtained for a number of quantities
and the free energy functional is still general enough to produce
for example a eutectic phase diagram.

\subsubsection{Solid-solid coexistence}

Expanding $F_{sol}$ around $\dN=0$ gives,
\be
F_{sol} =  F_{sol}(0) + a\, \dN^2/2 + b\, \dN^4/4 + \cdots
\ee 
where
\ben 
a &\equiv& w +  3\,B^{\ell}_2\,(A^o_{min})^2/8 \nline b
&\equiv& u - \frac{6(B^{\ell}_2)^2\,A^o_{min}}{(15vA^o_{min}-4t)}
\een 
and $A^o_{min}\equiv A_{min}(\dN=0)$ (which is thus a function of
$\Delta B_0$). This simple form can be used to calculate the
solid/solid coexistence concentrations at low temperatures according
to
\be 
\dN_{coex} = \pm \sqrt{|a|/b}. \label{eq:coexss}
\ee 
The critical temperature, $\Delta B_0^C$  is determined by setting
$\dN_{coex}=0$ and solving for $\Delta B_0$, which gives,
\be 
\Delta B^c_0 =
\left(15wv-2t\sqrt{-6B^{\ell}_2w}\right)/\left(6B^{\ell}_2\right).
\ee

\subsubsection{Solid-liquid coexistence}

    To obtain the liquid/solid coexistence lines the free energy of the
liquid state must be compared to that of the solid.   The mean
field free energy of the liquid state is obtained by setting
$n=0$ which gives,
\be 
 F_{liq} = \frac{w}{2} \dN^2 + \frac{u}{4} \dN^4.
\ee 

    To obtain the solid-liquid coexistence lines it is useful to expand the
free energy of the liquid and solid states around the value of $\dN$
at which the liquid and solid states have the same free energy,
i.e., when $F_{sol}=F_{liq}$. This occurs when,
\ben
\dN_{ls} &=& \pm\sqrt{\left(\Delta B_0^{ls}-\Delta
B_o\right)/B^{\ell}_2},
\een
where $\Delta B_o\equiv B_0^l-B_0^s$ and $\Delta B_0^{ls}\equiv
8t^2/(135v)$ is the lowest value of $\Delta B_0^{ls}$ at which a liquid
can coexist with a solid. To complete the calculations, $F_{sol}$
and $F_{liq}$ are expanded around $\delta N_{ls}$ to order $(\delta
N_{ls})^2$, i.e.,
\ben
\label{eq:freels}
 F_{liq} &=& {\cal F}(\dN_{ls}) + a_{liq}(\dN-\dN_{ls})
+ b_{liq}(\dN-\dN_{ls})^2 \cdots \nline
 F_{sol} &=& {\cal F}(N_{ls}) + a_{sol}(\dN-\dN_{ls})
+ b_{sol}(\dN-\dN_{ls})^2 \cdots \nline
\een
(note by definition
$F_{sol}(\dN_{ls}) = F_{liq}(\dN_{ls})$), where
\ben
a_{liq} &=& (w+u\dN_{ls}^2)\dN_{ls}  \nline a_{sol} &=&
a_{liq}+32t^2B^{\ell}_2\dN_{ls}/(675v^2)  \nline b_{liq} &=&
(w+3u\dN_{ls}^2)/2   \nline b_{sol} &=&
b_{liq}+\frac{6}{5v}B^\ell_2\left(\frac{4}{3}\Delta B_0 -\Delta
B^{ls}_0\right)
\een
It is now straightforward to calculate the liquid/solid coexistence
lines from Eq. (\ref{eq:freels}) in terms of the parameters given
above. The liquid/solidus lines are; \ben \dN_{liq} &=&
\dN_{ls}+\frac{1}{2} \lt(\frac{a_{liq}-a_{sol}}{b_{sol}-b_{liq}}\rt)
\lt(1-\sqrt{\frac{b_{sol}}{b_{liq}}}\rt) \nline \dN_{sol} &=&
\dN_{ls}+\frac{1}{2} \lt(\frac{a_{liq}-a_{sol}}{b_{sol}-b_{liq}}\rt)
\lt(1-\sqrt{\frac{b_{liq}}{b_{sol}}}\rt). \label{eq:coexls} \een for
$\dN_{liq} >0$, $\dN_{sol}>0$ and similar results for $\dN_{liq}
<0$, $\dN_{sol}<0$, since $F$ is a function of $\dN^2$ in this
example. The calculations in this section and the previous section
are reasonably accurate when $\Delta B_0^{ls} > \Delta B_0^c$,
however in the opposite limit a eutectic phase diagram forms and the
accuracy of the calculations decreases. This case will be discussed
below.

\subsubsection{Linear clastic constants}

    As shown in previous publications \cite{eg04,ekhg02},
the elastic constants can be calculated analytically in a one mode
approximation by considering changes in $F$ as a function of strain.
For the binary model similar calculations can be made in a
constant $\dN$ approximation and give,
\be
C_{11}/3=C_{12}=C_{44} = \frac{3}{16} B^s (A_{min})^2.
\ee
(this calculation can be done for arbitrary
dependence of $B^s$ on $\dN$).
As expected the elastic constants are directly proportional
to the amplitude of the density fluctuations.  This implies
that the elastic constants decrease as the liquid solid transition
is approached from the solid phase.  This result implies both a
temperature and concentration dependence through the
dependence of $A_{min}$ on $\Delta B$.  In addition to this
dependence (which might be considered as a 'thermodynamic'
dependence) the magnitude of the elastic constants can be
altered by the constant $B^s$.

\subsubsection{Calculation of phase diagrams}

    To examine the validity of some of the approximations made in the
previous section, numerical simulations were conducted to determine
the properties of the solid and liquid equilibrium states.  The
simulations were performed over a range of $\dN$ values, three
values of $\Delta B_0$ and two values of $w$.  The specific values
of the constants that enter the model are given in the figure
captions.

\begin{figure}[btp]
\center{\includegraphics*[width=0.40\textwidth]{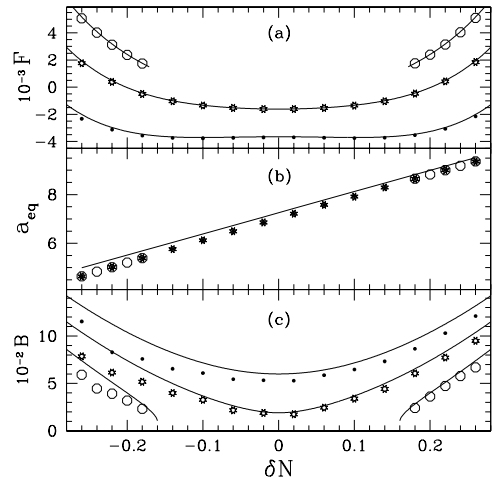}}
\caption{Free energy (a), lattice constant (b) and bulk modulus (c)
are plotted as a function of $\delta N$ for three different values
of $\Delta B_0$. In figure (a) the lines from top to bottom (and
bottom to top in (c)) correspond to $\Delta B_0=0.07$, $0.02$ and
$-0.03$ respectively computed in the one-mode approximations.  The
solid, starred and open points correspond to $\Delta B_0 = 0.07$,
$0.02$ and $-0.03$ and were calculated directly by numerically by
minimizing the free energy functional given in Eq. (\ref{eq:free}).
Other parameters used correspond $B^s_0=1.00$, $B^\ell_1=0$,
$B^\ell_2=-1.80$, $t=0.60$, $v=1.00$, $w=0.088$, $u=4.00$, $L=4.00$,
$Ro=1.00$ and $R_1=1.20$ (see Eq.~\ref{BR_exp} for definitions of
$R_o$ and $R_1$). } \label{fig:WFB}
\end{figure}

\begin{figure}[btp]
\center{\includegraphics[width=0.40\textwidth]{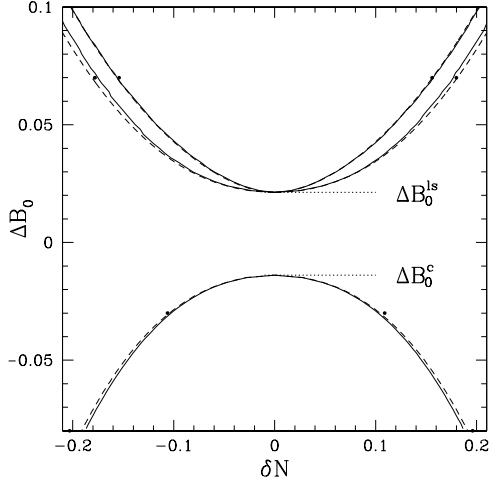}}
\caption{Phase diagram of $\Delta B_0$ Vs. $\delta N$ for the
parameters corresponding to the parameters of Fig. (\ref{fig:WFB}).
The solid line is a numerical solution of the one mode approximation
and the dashed lines are from Eq. (\ref{eq:coexss}) for the lower
solid/solid coexistence lines and Eq. (\ref{eq:coexls}) for the
upper liquid/solid coexistence lines. The solid points are from
numerical solutions for the minimum free energy functional given in
Eq. (\ref{eq:free}). }
\label{fig:free16}
\end{figure}

    In Fig. (\ref{fig:WFB}) analytic predictions and numerical solutions
are shown for the free energy, $F$, the lattice constant $R$ and
bulk modulus at three values of $\Delta B_0$, using the free energy
functional given in Eq. (\ref{eq:free}).  These figures indicate
that the approximate solutions are quite accurate except for some
deviations in the $\delta N$ dependence of the bulk modulus. For the
specific set of parameters used for these comparisons, the one-mode
constant concentration approximation predicts no asymmetry in any
quantity. However, it is clear from the numerical solutions that
some symmetry does exist. Figure (\ref{fig:free16}) shows the phase
diagram corresponding to the parameters used in Fig.~(\ref{fig:WFB})

\begin{figure}[btp]
\center{\includegraphics[width=0.40\textwidth]{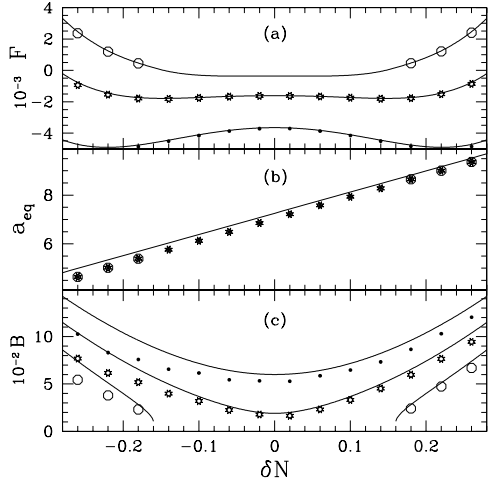}}
\caption{This figure is identical to Fig. (\ref{fig:WFB})
with the exception that $w=-0.04$.}
\label{fig:WFBE}
\end{figure}

\begin{figure}[btp]
\center{\includegraphics[width=0.40\textwidth]{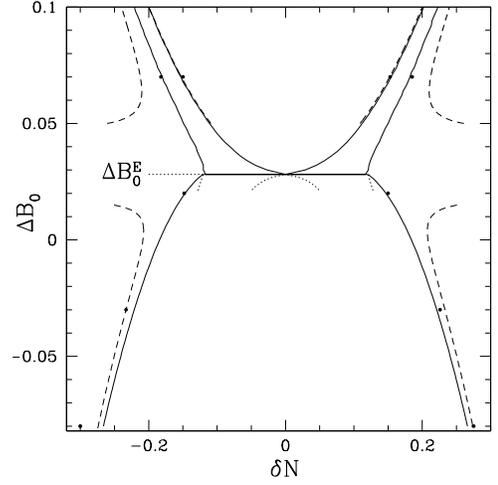}}
\caption{Phase diagram of $\Delta B_0$ Vs. $\delta N$ for the same
parameters as those used to generate Fig.~(\ref{fig:free16}), with
the exception that $w=-0.04$. The dotted lines below the eutectic
temperature, $\Delta B_0^E \approx 0.028$, correspond to metastable
states. }
\label{fig:free08}
\end{figure}

    The same calculations presented in Figs. (\ref{fig:WFB}) and
(\ref{fig:free16}) were repeated using $w=-0.04$ for which $\Delta
B_0^c > \Delta B_0^{ls}$. In this case the agreement between the
numerical results and the one-mode constant concentration
calculations for the free energy, lattice constant and bulk modulus
are similar to the $w=0.088$ case as shown in Fig. (\ref{fig:WFBE}).
The phase diagram corresponding to the parameters used in
Fig.~(\ref{fig:WFBE}) is shown in Fig.~(\ref{fig:free08}). In this
case, the analytic calculations (Eqns. (\ref{eq:coexls}) and
(\ref{eq:coexss})) for the coexistence lines breakdown at the
eutectic point. Higher order terms in the in $\delta N$ are needed
to accurately predict the coexistence lines.

\section{Dynamics}
\label{EOMS}

    To simulate microstructure formation in binary alloys, dynamical
equations of motions for the field $\delta N$ and $n$ need to be
developed. The starting point is the full free energy in
Eq.~(\ref{full_free_eng}), written in terms of in terms of $\rho_A$
and $\rho_B$, i.e., ${\cal F}(\rho_A,\rho_B)$. The dynamics of
$\rho_A$ and $\rho_B$ is assumed to be dissipative and driven by
free energy minimization, i.e.,
\ben 
\frac{\partial \rho_A}{\partial t}=\nabla \cdot \lt(M_A
\lt(\rho_A,\rho_B \rt)
\nabla \frac{\delta {\cal F}}{\delta \rho_A} \rt) \\
\frac{\partial \rho_B}{\partial t}=\nabla \cdot \lt(M_B
\lt(\rho_A,\rho_B \rt) \nabla \frac{\delta {\cal F}}{\delta \rho_B}
\rt)
\een 
where $M_A$ and $M_B$ are the mobilities of each density field. In
general these mobilities depend on the density of each species. The
free energy ${\cal F}(\rho_A,\rho_B)$ can equivalently be defined in
terms of $n$ and $\delta N$. This allows the previous equations to
be re-written as
\ben
\bar{\rho} \frac{\partial \rho_A}{\partial t}=\lt( M_A \nabla^2 +
\nabla \cdot M_A  \nabla \rt) \lt( \frac{\delta {\cal F}}{\delta n}
- \frac{\delta {\cal F}}{\delta \bcc} \rt)
\label{eq3} \\
\bar{\rho} \frac{\partial \rho_B}{\partial t}=\lt( M_B \nabla^2 +
\nabla \cdot M_B \nabla \rt) \lt( \frac{\delta {\cal F}}{\delta n} +
\frac{\delta {\cal F}}{\delta \bcc} \rt) \label{eq4}.
\een 

    Adding Eq.(\ref{eq3}) and Eq.(\ref{eq4}) and collecting terms gives
the following equation for $n$,
\begin{align} 
\frac{\partial n}{\partial t} = \lt\{ M_1 \nabla^2 + \nabla M_1
\cdot \nabla \rt\} \frac{\delta {\cal F}}{\delta n} \nonumber \\
+ \lt\{ M_2 \nabla^2 +  \nabla M_2 \cdot \nabla \rt\} \frac{\delta
{\cal F}}{\delta \bcc}. \label{n-equation}
\end{align} 
where $M_1 \equiv (M_A+M_B)/\bar{\rho}^2$ and $M_2
\equiv (M_B - M_A)/\bar{\rho}^2$. Subtracting Eq.(\ref{eq3}) from
Eq.(\ref{eq4}) similarly gives an equation for  $\delta N$,
\begin{align} 
\frac{\partial \bcc}{\partial t}= \lt\{ M_2 \nabla^2 + \nabla M_2
\cdot \nabla \rt\} \frac{\delta {\cal F}}{\delta n} \nonumber \\
+\lt\{ M_1 \nabla^2 + \nabla M_1 \cdot \nabla \rt\} \frac{\delta
{\cal F}}{\delta \bcc}. \label{c-equation}
\end{align} 
Equations (\ref{n-equation}) and (\ref{c-equation}) can be cast into
the more illuminating form
\begin{align} 
\frac{\partial n}{\partial t}= \vec{\nabla}\cdot \lt\{ M_1
\vec{\nabla} \frac{\delta {\cal F}}{\delta n}\rt\} +
\vec{\nabla}\cdot\lt\{ M_2 \vec{\nabla}
\frac{\delta {\cal F}}{\delta \bcc} \rt\}  \label{final_n_equation}\\
\frac{\partial \bcc}{\partial t}= \vec{\nabla}\cdot \lt\{ M_2
\vec{\nabla} \frac{\delta {\cal F}}{\delta n}\rt\} +
\vec{\nabla}\cdot\lt\{ M_1 \vec{\nabla} \frac{\delta {\cal
F}}{\delta \bcc} \rt\} \label{final_c_equation}.
\end{align} 
Equations (\ref{final_n_equation}) and (\ref{final_c_equation})
couple the dynamics of the fields $\delta N$ and $n$ through a
symmetric mobility tensor. The dependence of the mobilities $M_A$
and $M_B$ will in general depend on local crystal density and the
local relative concentration of species $A$ to $B$.

    For the case of substitutional diffusion between species $A$
and $B$, $M_A \approx M_B \equiv M$. In this limit
the dynamics of $n$ and $\delta N$ decouple. Moreover if
it is further assumed that that the mobility is a constant,
Eqs.(~\ref{final_n_equation}) and (\ref{final_c_equation}) become
\be 
\frac{\partial n}{\partial t}= M_e \nabla^2 \frac{\delta {\cal
F}}{\delta n} \label{simple_n_eqn}
\ee 
\be 
\frac{\partial \bcc}{\partial t}=M_e \nabla^2 \frac{\delta {\cal
F}}{\delta \bcc} \label{simple_c_eqn}
\ee 
where the effective mobility $M_e \equiv 2M/\bar{\rho}^2$. In the
applications using Eqs.~(\ref{simple_n_eqn}) and
(\ref{simple_c_eqn}) in the following sections, the dynamics of $n$
and $\delta N$ are simulated with time re-scaled by $t \rightarrow
\bar{t} \equiv 2M t/\bar{\rho}^2$.

\begin{figure}[btp]
\vskip -1.5cm
\center{\includegraphics[width=0.50\textwidth]{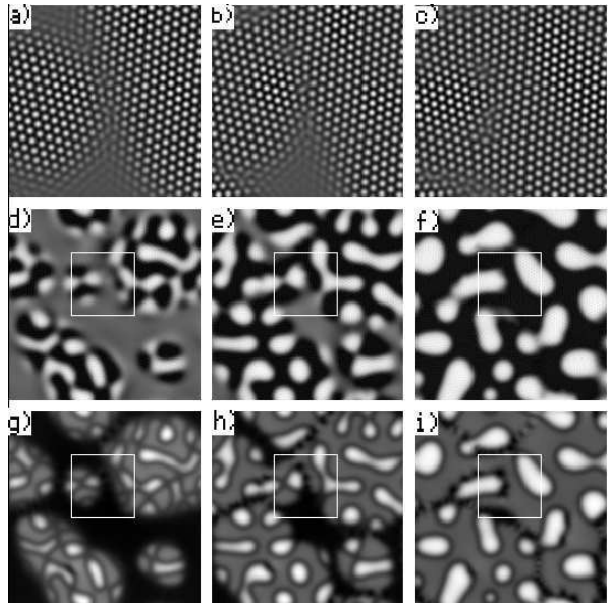}}
\vskip -1cm \caption{The grey scales in the figure correspond to
density ($n$), concentration ($\delta N$) and the local energy
density in frames (a,b,c), (d,e,f) and (g,h,i) respectively.  The
area enclosed by white boxes is area shown in figures (a,b,c).  The
parameters in this simulation are the same as Fig.
(\ref{fig:free08}) except $L=1.20$ and $R_1/R_0 = 1/4$ at $\delta N
=0$ and $\Delta B_o = 0.02$. Figures (a,d,g), (b,e,h) and (c,f,i)
correspond to times $t = 6600, 16200, 49900$, respectively. }
\label{fig:allstuff}
\end{figure}

    To illustrate the dynamics described by Eqs.~(\ref{simple_n_eqn}),
(\ref{simple_c_eqn}) and (\ref{eq:free}), a simulation of
heterogenous eutectic crystallization from a supercooled homogenous
liquid was performed. The results of this simulation are shown in
Fig. (\ref{fig:allstuff}).  This figure shows the density ($n$), the
density difference ($\delta N$) and the local energy density at
three time steps in the solidification process.   These figures show
liquid/crystal interfaces, grain boundaries, phase segregation,
dislocations and multiple crystal orientations all in a single
numerical simulation of the simple binary alloy PFC model.  In this
simulation a simple Euler algorithm was used for the time derivative
and the spherical Laplacian approximation introduced was used (see
Appendix A). The grid size was $\Delta x = 1.1$ and the time step
was $\Delta t = 0.05$. Unless otherwise specified all simulations to
follow use the same algorithm, grid size and time step.

\section{Applications}

This section applies the simplified phase field crystal model
derived in Section \ref{sec:simple}, coupled to the dynamical
equations of motion derived in Section \ref{EOMS}, to the study of
elastic and plastic effects in phase transformations. The first
application demonstrates how the PFC alloy model can be used to
simulate eutectic and dendritic microstructures. That is followed by
a discussion of the effects of compressive and tensile stresses in
epitaxial growth. Finally, simulations demonstrating the role of
dislocations in Spinodal decomposition are presented.

\subsection{Eutectic and Dendritic Solidification}

    One of the most important applications of the alloy phase field
crystal model is the study of solidification microstructures. These
play a prominent role in numerous applications such as commercial
casting. Traditional phase field models of solidification are
typically unable to self-consistently combine bulk elastic and
plastic effects with phase transformation kinetics, multiple crystal
orientations and surface tension anisotropy. While some of these
effects have been included in previous approaches (e.g. surface
tension anisotropy) they are usually introduced phenomenologically.
In the PFC formalism, these features arise naturally from density
functional theory.

\begin{figure}[btp]
\center{\includegraphics[width=0.40\textwidth]{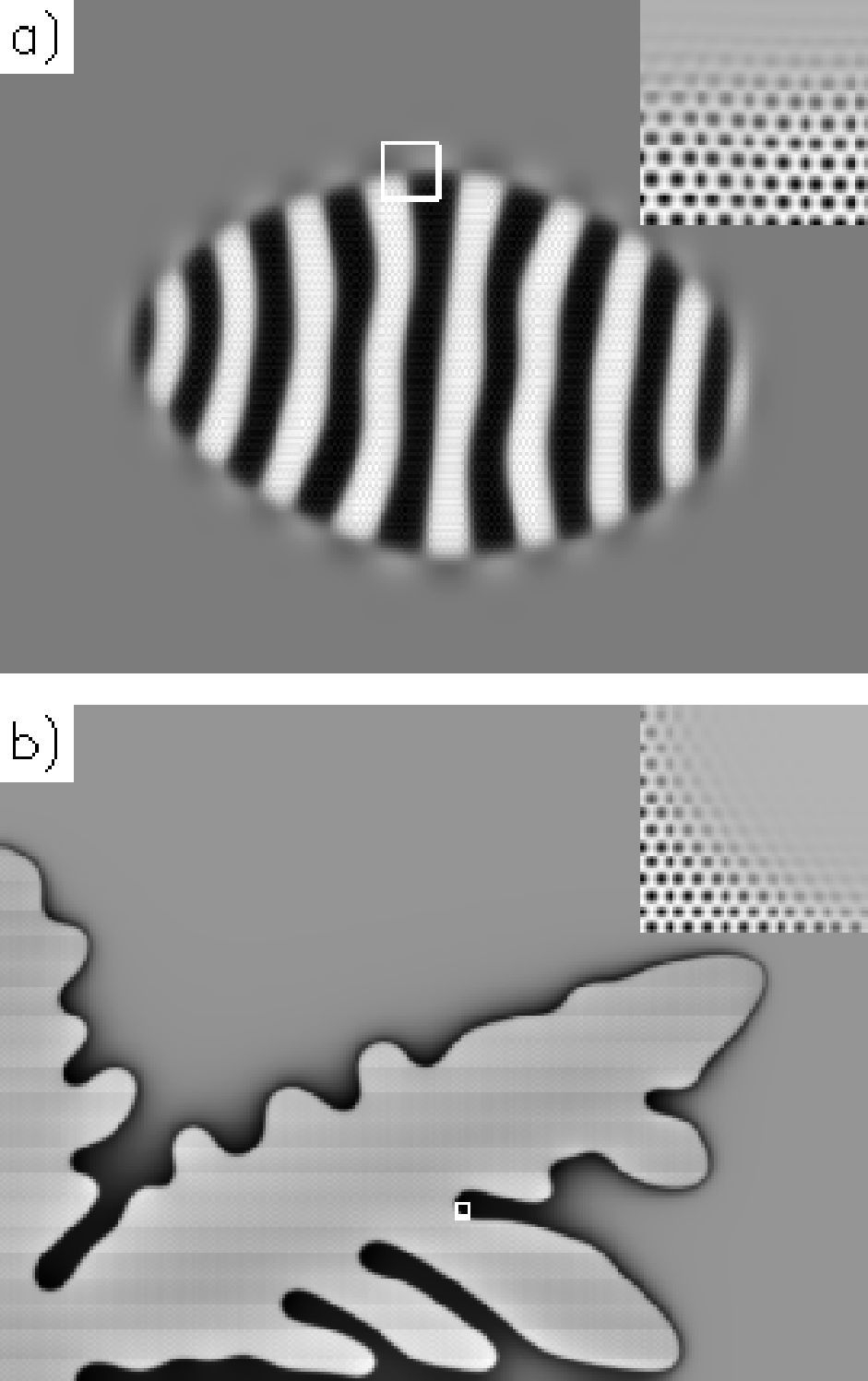}}
\caption{The grey-scale in the main portion of both figures show the
concentration field $\dN$. In the insets, the grey-scale shows the
density field, $n$, for the small portion of the main figure that is
indicated by the white boxes. a) Eutectic crystal grown from a
supercooled liquid at $\Delta B_0 = 0.0248$ and $\dN = 0.0$. The
parameters that enter the model are the same as Fig.
(\ref{fig:free08}) except $L=1.20$ and $R_1/R_0 = 1/4$. b) Dendrite
crystal grown from a supercooled liquid at $\Delta B_0 = 0.04$ and
$\dN = 0.0904$.   The parameters that enter the model are the same
as Fig. (\ref{fig:free16}) except $L=1.20$ and $R_1/R_0 = 1/4$. In
Fig. (b) mirror boundary conditions were used.} \label{fig:eutden}
\end{figure}

    To illustrate solidification microstructure formation using the
PFC formalism, two simulations were conducted that the growth of a
single crystal from a supercooled melt in two dimensions. In the
first simulation a small perturbation in the density field was
introduced into a supercooled liquid using the parameters
corresponding to the phase diagram in Fig. (\ref{fig:free08}),
except $L=1.20$ and $R_1/R_0 =1/4$. The reduced temperature $\Delta
B_0 = 0.0248$ and average concentration of $\dN= 0.0$.  To reduce
computational time the size of the lattice was gradually increased
as the seed increased in size. A snapshot of the seed is shown at
$t=480,000$ in Fig. (\ref{fig:eutden}a). A similar simulation was
conducted for the growth of a dendrite from a supercooled melt for
reduced temperature $\Delta B_0 = 0.04$ and $\dN = 0.0904$, with
other parameters corresponding to those in Fig. (\ref{fig:free16}),
except for $L=1.20$ and $R_1/R_0 = 1/4$. A sample dendritic
structure is shown in Fig. (\ref{fig:eutden}b) at $t=175,000$.

    Simulations such as these can play an important role in
establishing various constitutive relations for use in higher-scale
finite element modeling (FEM) of elasto-plastic effects in alloys
during deformation. In particular, traditional FEM approaches often
employ empirical or experimental constitutive models to describe
stress-strain response in elements that are intended to represent
one (or more) grains. These constitutive relations are often limited
in their usefulness as they do not self-consistently incorporate
realistic information about microstructural properties that develope
during solidification.

\subsection{Epitaxial Growth}

    Another potential application of the PFC model is in the
technologically important process of thin film growth. The case of
heteroepitaxy, the growth of a crystalline film exhibiting atomic
coherency with a crystalline substrate of differing lattice constant,
has been examined in previous PFC studies of pure systems
\cite{eg04,ekhg02}. These initial works focused on two of the
primary phenomena influencing film quality: (i) morphological
instability to buckling or roughening and (ii) dislocation nucleation
at the film surface. A third important effect in alloy films, (iii)
compositional instability (phase separation in the growing film),
requires consideration of multiple atomic species and their interaction.
The purpose of this section is to illustrate how the binary PFC model
addresses such compositional effects in alloy heteroepitaxy, focusing
on the spatial dynamics of phase separation over diffusive time scales.

    To date, a number of models of single component film growth
incorporating surface roughening, dislocation nucleation, or both
have been proposed
\cite{mb74,at72,g82,pb85,mg99,hmrg02,ys93,kmm01,tl94,gn99}, and
models of binary film growth incorporating surface roughening and phase
separation have been proposed as well \cite{svt00,gv96,hd02,ld978}.
However, no existing
models of binary film growth known to the authors have captured all
of the above important phenomena, and it would be reasonable to
expect that new insights into the nature of film growth could be
gained through the simultaneous investigation of all of these growth
characteristics. A unified treatment of this sort is required for
the following reasons. There is clearly a strong link between surface
roughening and dislocation nucleation, originating from the fact
that dislocations nucleate at surface cusps when the film becomes
sufficiently rough. It is also known that phase separation in the
film is significantly influenced by local stresses, which are
inherently coupled to surface morphology and dislocation nucleation.
The dynamics of the growth process must then be influenced by the
cooperative evolution of all three of these phenomena.  In the next
subsection numerical simulations will be presented to show that the
binary PFC model produces all of the growth characteristics described
above, and that each is influenced by misfit strain and atomic size
and mobility differences between species.

\subsubsection{Numerical simulations}

    The physical problem recreated in these simulations is that of growth
of a symmetric (50/50 mixture, $\delta N_0 =0$) binary alloy film from a
liquid phase or from a saturated vapor phase above the bulk coherent
spinodal temperature ($\Delta T_c$).  Growth at temperatures above
the miscibility gap is typical of experimental conditions and should
ensure that phase separation is driven by local stresses and is not
due to spinodal decomposition.  Initial conditions consisted of a
binary, unstrained crystalline substrate, eight atoms in thickness,
placed below a symmetric supercooled liquid of components $A$ and
$B$.  In all the simulations presented, parameters are the same as in Fig.
(\ref{fig:WFB}) except for $L=1.882$ and $\Delta B = 0.00886$ unless
specified in the figure caption.  In what follows the misfit strain,
$\epsilon$, is defined as $(a_{film} - a_{sub})/a_{sub}$, where
$a_{film} \approx a_A(1+\eta \delta N_0)$ if in the constant concentration
approximation.  For a symmetric mixture of $A$ and $B$ atoms (i.e.,
$\delta N_0=0$) $a_{film} = (a_A + a_B)/2$.

    Periodic boundary conditions were used in the lateral directions,
while a mirror boundary condition was applied at the bottom of the
substrate. A constant flux boundary condition was maintained along
the top boundary, $120\Delta x$ above the film surface, to simulate
a finite deposition rate. Misfit strain was applied to the system by
setting $R=1$ in the substrate and $R=1+\epsilon+\eta \delta N$
in the film.  This approach yields a film and substrate that are
essentially identical in nature except for this shift in lattice
parameter in the film.  Complexities resulting from differing
material properties between the film and substrate are therefore
eliminated, isolating the effects of misfit strain, solute strain,
and mobility differences on the film growth morphology.  The
substrate was permitted to strain elastically, but was prevented
from decomposing compositionally except near the film/substrate
interface.

\begin{figure}[btp]
\center{\includegraphics[width=0.40\textwidth]{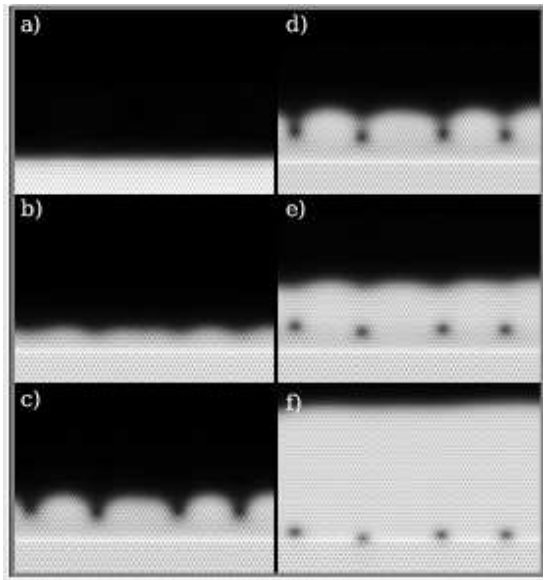}}
\caption{Plots of the smoothed local free energy showing the
progression of the buckling instability, dislocation nucleation and
climb towards the film/substrate interface.  From a) to f) times
shown are t=150, 600, 1050, 1200, 1500 and 2550. In this figure
$\epsilon=0.04$, $\eta=-1/4$ and $M_A=M_B=1$.} \label{fig:atgprog1}
\end{figure}

    A sample simulation is shown in Fig. (\ref{fig:atgprog1}) for a
misfit strain of $\epsilon=0.04$, a solute expansion coefficient of
$\eta=-1/4$ and mobilities $M_A=M_B=1$.   As seen in this
figure the well-documented buckling or Asaro-Tiller-Grinfeld
\cite{at72,g82} instability is naturally reproduced by the PFC model.
This instability is ultimately suppressed as a cusp-like surface morphology
is approached, with increasingly greater stress developing
in surface valleys.  The buckling behavior ceases only when the local
stress in a given valley imparts on the film a greater energy than that
possessed by an equivalent film with a dislocation.  At this stage a
dislocation is nucleated in the surface valley and the film surface begins
to approach a planar morphology.

\begin{figure}[btp]
\center{\includegraphics[width=0.40\textwidth]{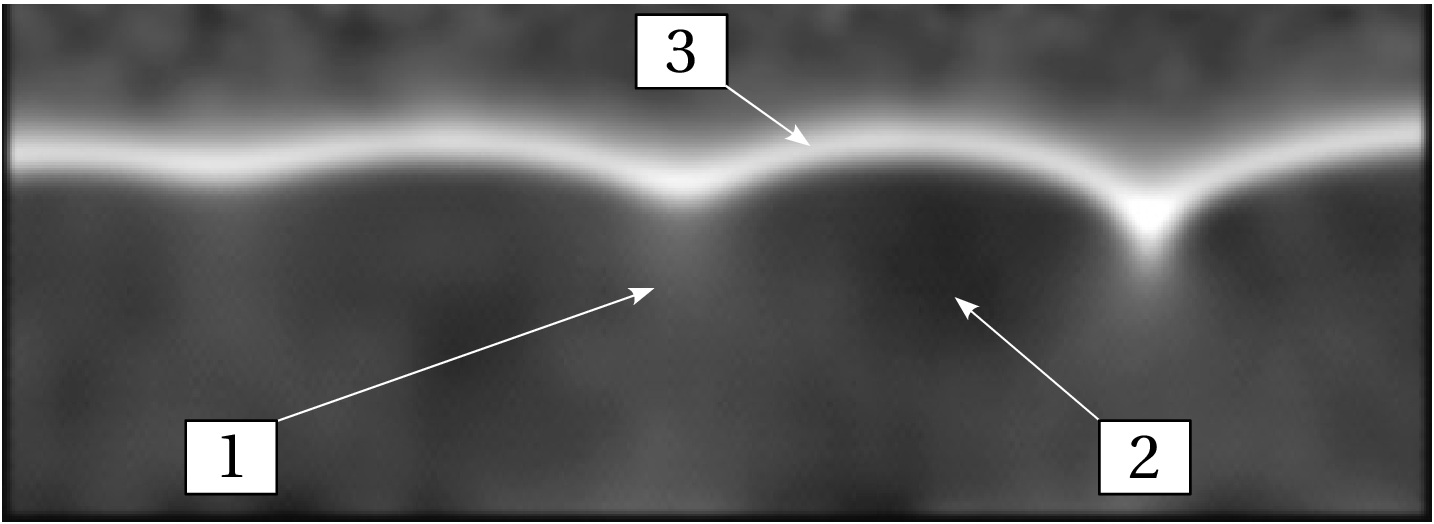}}
\caption{Plot of the smoothed local concentration field showing
lateral phase separation between the surface peaks and valleys.
White: Component A (large, fast), Black: Component B (small, slow).
In this figure $\epsilon=-0.02$,
$\eta=0.4$, $M_A=1$, $M_B=1/4$, and $t=3500$. See text for discussion
of the numbered arrows.} \label{fig:mainc}
\end{figure}

\begin{figure}[btp]
\center{\includegraphics[width=0.40\textwidth]{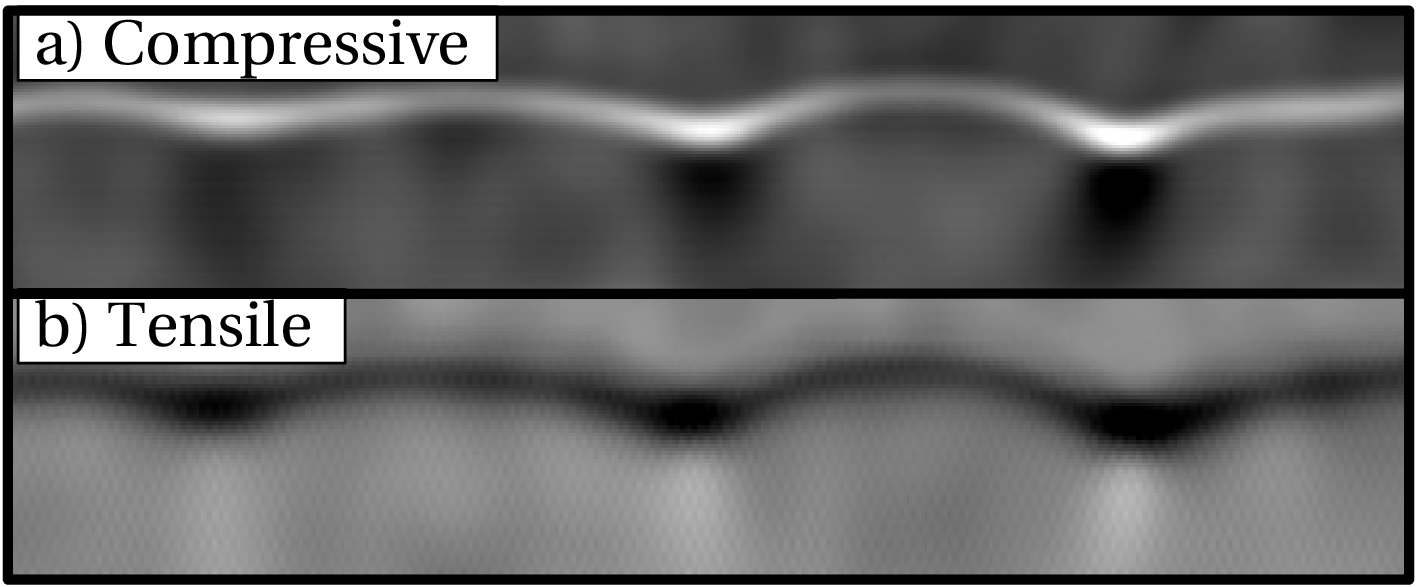}}
\caption{Plot of the smoothed local concentration field showing the
nature of the phase separation under opposite signs of $\epsilon$.
In figures a) and b) $\epsilon=0.04$ and $-0.025$ respectively and
in both figures $M_A=M_B=1$ and $\eta=0.25$.} \label{fig:comptensc}
\end{figure}

    The nature of phase separation within the bulk film and at the film
surface was found to vary with model parameters, but a number of
generalizations applicable to all systems studied have been identified.
For the case of equal mobilities ($M_A=M_B$) we find that in the
presence of misfit and solute strain, the component with greater
misfit relative to the substrate preferentially segregates
below surface peaks (see regions marked $1$ and $2$ in Fig.
(\ref{fig:mainc}) and Fig. (\ref{fig:comptensc})). Larger (smaller)
atoms will be driven toward regions of tensile (compressive) stress
which corresponds to peaks (valleys) in a compressively strained
film and to valleys (peaks) in a film under tensile strain. This
coupling creates a lateral phase separation on the length scale of the
surface instability and has been predicted and verified for binary films
\cite{svt00,gv96,hd02,ld978,walther97,okada97,peiro95,mmt97,pbl97}
and analogous behavior has been predicted and verified in quantum dot
structures \cite{ltb00,tersoff98,rrl04}.

    Secondly, again for the case of equal mobilities, the component with
greater misfit relative to the substrate is driven toward the film surface
(see Fig. (\ref{fig:comptensc})).  This behavior can also be explained in
terms of stress relaxation and is somewhat analogous to impurity rejection
in directional solidification.  The greater misfit component can be viewed
as an impurity that the growing film wishes to drive out toward the
interface. Experimental evidence from SiGe on Si \cite{walther97} and
InGaAs on InP \cite{okada97,peiro95}
verifies this behavior as an enrichment of the greater misfit component
was detected at the film surface in both systems. Other models
\cite{svt00,gv96,hd02,ld978} have
not led to this type of vertical phase separation possibly due to
neglecting diffusion in the bulk films.

\begin{figure}[btp]
\center{\includegraphics[width=0.40\textwidth]{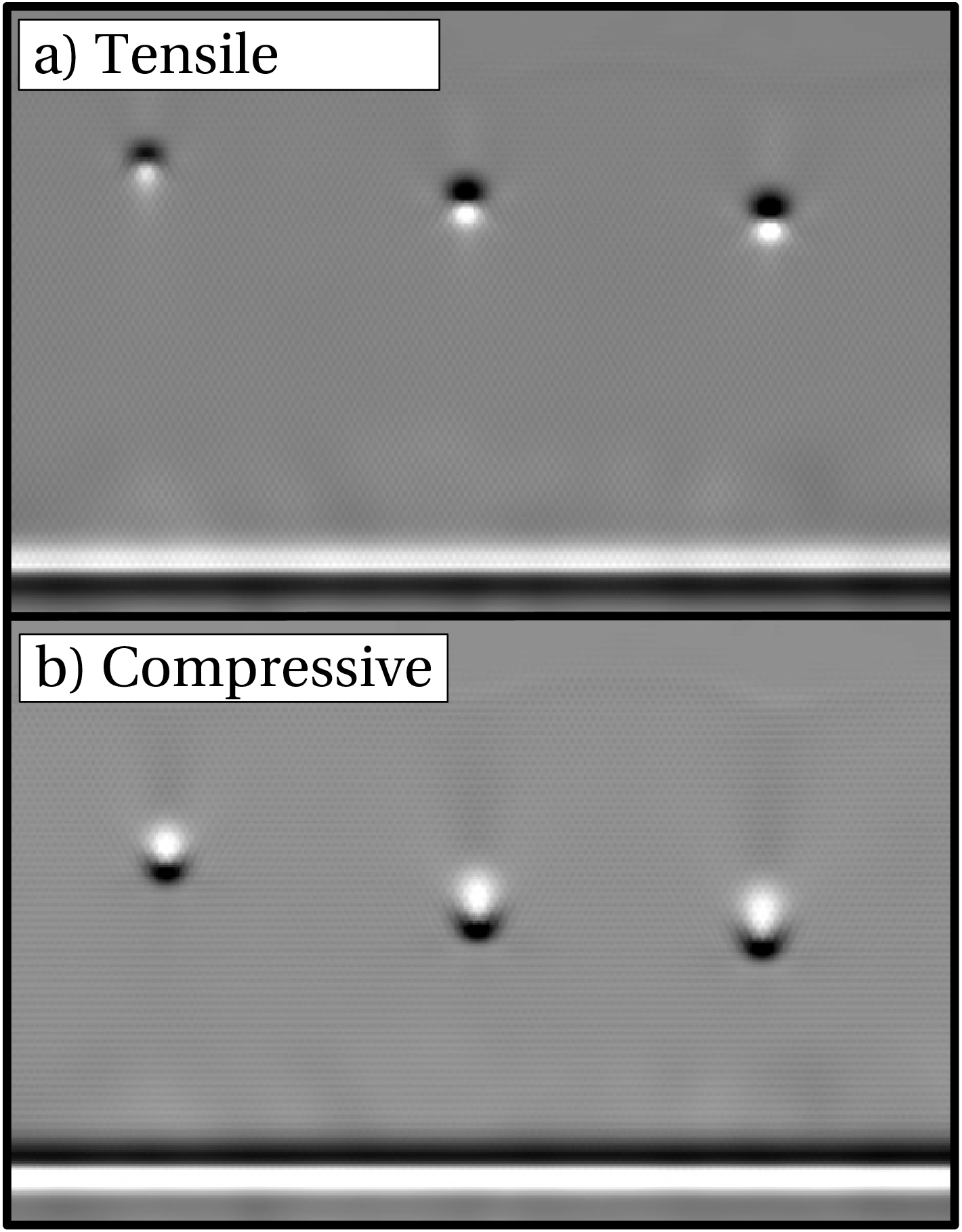}}
\caption{Plot of the smoothed local concentration field showing the
complimentary phase separation at the film/substrate interface and
around defects. In this figure the film/liquid interface cannot be
distinguished due to the overwhelming contrast created near the
defects and at the film/substrate interface. In figures a) and b)
$\epsilon=-0.035$ and $0.05$ respectively, and in both figures
$M_A=M_B=1$ and $\eta=1/4$.} \label{fig:defc}
\end{figure}

    The third generalization that can be made is that, in the case of
sufficiently unequal mobilities, the component with greater mobility
accumulates at the film surface (see region marked $3$ in Fig.
(\ref{fig:mainc})). It was found that when
the two components have a significant mobility difference
(typically greater than a 2:1 ratio) the effect of mobility
is more important than the combined effects of misfit and solute
strains in determining which component accumulates at the surface.
Since Ge is believed to be the more mobile component in the SiGe
system, we see that the findings of Walther et al \cite{walther97} for
SiGe on Si provide experimental support for this claim.  They find a
significant enrichment of Ge at the film surface, a result that
was likely due to a combination of this mobility driven effect as
well as the misfit driven effect described in the second generalization.
Experimental evidence also indicates that segregation of
substrate constituents into the film may occur during film growth
\cite{jsb97,klw93}.  We have similarly found that a vertical
phase separation can be produced near the film/substrate interface and
is complimented by a phase separation mirrored in direction near
defects (see Fig. (\ref{fig:defc})).  The extent of this phase
separation is controlled largely by the bulk mobilities of
the two constituents, and to a lesser degree by $\eta$.
The complimenting phase separation near climbing defects
is a transient effect, any traces of which are dulled
once the defect reaches the film/substrate interface.

\subsection{The Role of Dislocations in Spinodal Decomposition}

    Spinodal decomposition is a non-equilibrium  process in which a
linearly unstable homogenous phase spontaneously decomposes into two
daughter phases.  An example of this process in the solid state
occurs during a quench below the spinodal in Fig.~(\ref{fig:free16})
when $\delta N=0$. During this process domains of alternating
concentration grow and coarsen to a tens of nanometers.  Spinodal
decomposition is of interest as it is a common mechanism for
strengthening alloys, due to the large number of interfaces that act
to impede dislocation motion.

    Solid state strengthening mechanisms, such as spinodal
decomposition, rely critically on the interactions that exist
between dislocations and phase boundaries. Cahn was first to
calculate that the driving force for nucleation of an incoherent
second phase precipitate is higher on a dislocation than in the bulk
solid \cite{Cah57}. A similar result was obtained by Dolins for a
coherent precipitate with isotropic elastic properties in the solid
solution \cite{Dol70}. Hu et. al confirmed the results of Cahn and
Dolins using a model that included elastic fields from compositional
inhomogeneities and structural defects \cite{Hu02}.

    Recent studies of spinodal decomposition have used phase field
models to examine the role of dislocations on alloy hardening
\cite{Cah63, Rod03}. These phase field models couple the effects of
static dislocations to the kinetics of phase separation. L\'{e}onard
and Desai where the first to simulate the effect of static
dislocations on phase boundaries, showing that the presence of
dislocations strongly favors the phase separation of alloy
components \cite{Leo98}.

    Haataja et al. recently introduced mobile dislocations into a phase
field model that couples two burgers vectors fields to solute
diffusion and elastic strain relaxation. It was shown that mobile
dislocations altered the early and intermediate time coarsening
regime in spinodal decomposition \cite{Haa04,Haa05}. Specifically,
it was found that coherent strain at phase boundaries decrease the
initial coarsening rate, since they increase stored elastic energy
in the system. As dislocations migrate toward moving interfaces,
they relax the excess the strain energy, thus increasing the
coarsening rate \cite{Haa05}. The growth regimes predicted by the
model in Ref.~\cite{Haa05} are in general agreement with several
experimental studies of deformation on spinodal age hardening
\cite{Bha96, Ple75, Hel77, Spo77}.

\subsubsection{Numerical Simulations}

\begin{figure}[btp]
\center{\includegraphics[width=0.50\textwidth]{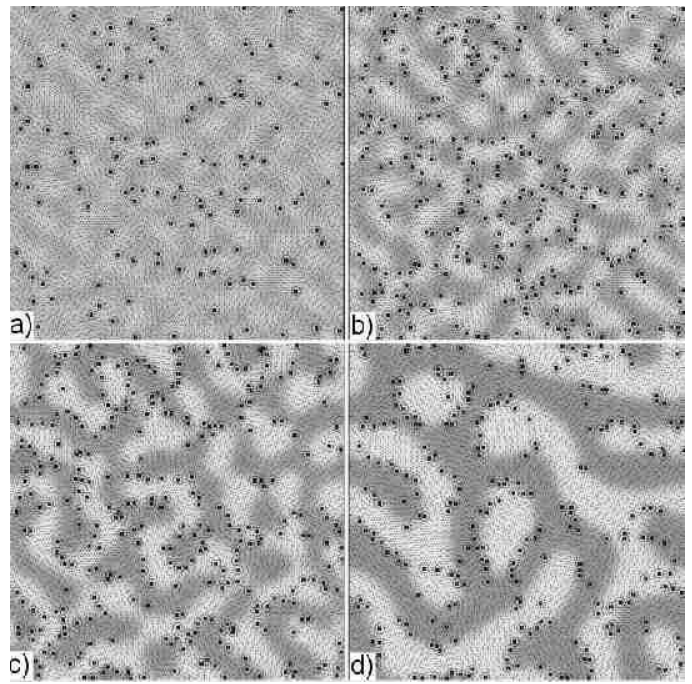}}
\caption{Four time sequences in the evolution of the concentration
field (grey scale), superimposed on the corresponding density field.
Dislocations are labeled by a square on the dislocation core
surrounded by a circle. The time sequence (a)-(d) corresponds to
$t=12000$, $24000$, $60000$ and $288000$, respectively (in units of
$\Delta t=0.004$). The system size is: $1024 \Delta x \times 1024
\Delta x$, where $\Delta x = \pi / 4$. The density difference
$\delta N=0$, while $L=2.65$, $R_1/R_0 = 0.25$ and all other
parameters are the same as Fig. (\ref{fig:free16}). }
\label{fig:SD1}
\end{figure}

\begin{figure}[btp]
\center{\includegraphics[width=0.40\textwidth]{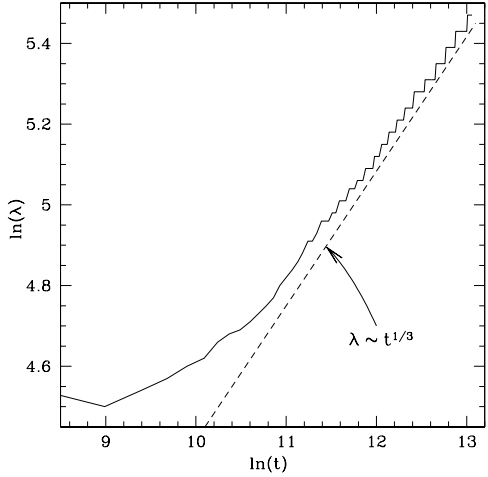}}
\caption{Inverse of the mean wave vector of the (circularly
averaged) 2D structure factor of the concentration field Vs. time
corresponding to the simulation in Figure ~\ref{fig:SD1}.}
\label{fig:SD2}
\end{figure}

    The findings of Ref.~\cite{Haa05} were compared with numerical
simulations from the alloy PFC model. Here, spinodal decomposition
was simulated for an alloy corresponding to the phase diagram in
Fig. (\ref{fig:free16}). Simulations began with a liquid phase of
average dimensionless density difference $\delta N=0$, which first
solidified into a polycrystalline solid (alpha) phase, which
subsequently phase separated as the reduced temperature ($\Delta
B_o$) was lowered below the spinodal. Figure~(\ref{fig:SD1}) shows
the concentration and density fields for four time sequences during
the spinodal decomposition process. The dots in the figures denote
the locations of dislocation cores. Parameters for this simulation
are given in the figure caption. Figure~(\ref{fig:SD2}) shows a plot
of the average domain size versus time corresponding to the data in
Figure~(\ref{fig:SD1}).

\begin{figure}[btp]
\center{\includegraphics[width=0.40\textwidth]{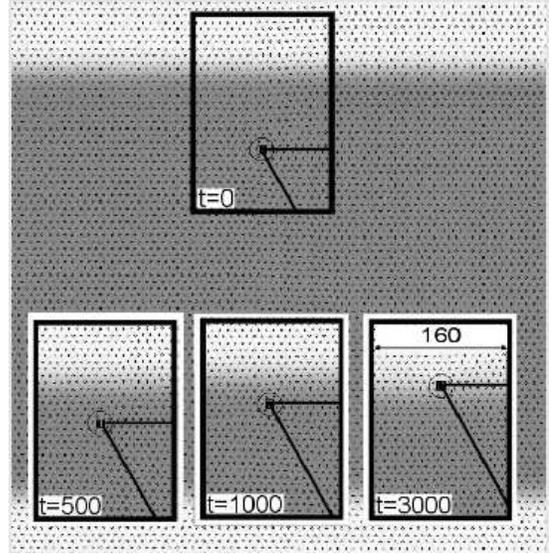}}
\caption{A dislocation migrates toward a coherent phase boundary
thus relaxing mismatch strain. An $800 \times 800$ (units of $\Delta
x$) portion of the actual simulations domain is shown. The data
shows four time frames in the motion of the dislocation. Parameters
of the simulation are the same as in Figure ~\ref{fig:SD1}.}
\label{fig:singled}
\end{figure}

    The PFC alloy results are in general agreement the results of
Ref.~\cite{Haa05}. Specifically, PFC simulations show an early and
intermediate time regime where the spinodal coarsening rate is is
reduced from its traditional $t^{1/3}$ behaviour. In this regime,
the strain energy stored in the system is found to be much higher
than that at late times when dislocations migrate to the boundaries,
relaxing strain energy and leading to an an increased coarsening
rate toward the usual $t^{1/3}$ growth law. To more clearly
illustrate the the interaction between a coherent boundaries and
dislocations, Fig.~(\ref{fig:singled}) shows four time steps in the
evolution of a dislocation migrating toward a phase boundary to
relax mismatch strains.

    It is noteworthy that the alloy PFC introduced in this work does not
incorporate ``instantaneous" elastic relaxation. A proper treatment
of rapid relaxation of strain fields requires the model to be
extended in a manner analogous to \cite{Ste06}. However, because of
the asymptotically slow kinetics of spinodal decomposition and the
small length scales between domain boundaries, it is expected that
this will only influence the time scales over which dislocations
interact with domain boundaries. As a result, the general trends
depicted in Figures.~(\ref{fig:SD2}) and (\ref{fig:SD1}) are
expected to be correct.

\section{Discussion and Conclusions}

    In this paper a connection between the density functional
theory of freezing and phase field modeling was examined. More
specifically it was shown that the phase field crystal model
introduced in earlier publications \cite{ekhg02,eg04,bme06} and the
regular solution commonly used in material science can be obtained
from DFT in certain limits.  These calculations relied on
parameterizing the direct two-point correlation function that enters
DFT by three quantities related to the elastic energy stored in the
liquid and crystalline phases, as well as the lattice constant.

    In addition, a simplified binary alloy model was developed that
self-consistently incorporates many physical features inaccessible
in other phase field approaches.  The simplified alloy PFC model was
shown to be able to simultaneously model solidification, phase
segregation, grain growth, elastic and plastic deformations in
anisotropic systems with multiple crystal orientations on diffusive
time scales.

It is expected that the alloy PFC formalism and its extensions can
play an important role in linking material properties to
microstructure development in a manner that {\it fundamentally}
links the meso-scale to the atomic scale.  As such this formalism,
particularly when combined with novel adaptive mesh techniques in
phase-amplitude or reciprocal space, can lead the way to a truly
multi-scale methodology for predictive modeling of materials
performance.

\section{Acknowledgments}
K. R. E. acknowledges support from the National Science Foundation
under Grant No. DMR-0413062.  N. P. and M. G. would like to thank
the National Science and Engineering Research Council of Canada for
financial support.  M. G. also acknowledges support from le Fonds
Quebecois de la recherche sur la nature et les technologies. J. B.
acknowledges support from the Richard H.  Tomlinson Foundation.

\appendix
\section{Numerical discretization Schemes}
\label{app:nummet}

    The governing Equations were numerically solved using two different
methods, described below.  In what follows the subscripts $n$, $i$
and $j$ are integers corresponding to the number of time steps and
distance along the $x$ and $y$ directions, respectively, of a
discrete numerical lattice. Time and space units are recovered by
the simple relations, $t=n\Delta t$, $x=i\Delta x$ and $y=j\Delta
x$.

    An Euler discretization scheme was used for the time derivative and
the `spherical laplacian' approximation was used to calculate all
Laplacians.  For this method the discrete dynamics become
\be \psi_{n+1,i,j} = \psi_{n,i,j}+\nabla^2 \mu_{n,i,j},
\ee
where $\mu_{n,i,j}$ is the chemical potential given by
\be
\mu_{n,i,j} = (r+(1+\nabla^2)^2)\psi_{n,i,j}-\psi_{n,i,j}^3.
\ee
All Laplacians were evaluated as follows,
\ben
&&\nabla^2f_{n,i,j}=\Big(\big(
f_{n,i+1,j}+f_{n,i-1,j}+f_{n,i,j+1}\nline
&+&f_{n,i,j-1}\big)/2+\big(f_{n,i+1,j+1}+f_{n,i-1,j+1}\nline
&+&f_{n,i-1,j+1}+f_{n,i-1,j+1}\big)/4-3f_{n,i,j}\Big)/(\Delta x)^2.
\een

\section{Free Energy Functional for a Pure Material}
\label{app:DFT}

    This appendix outlines the derivation of the free energy
functional ${\cal F}[\rho]$ presented in Section \ref{DFT:single}
for a non-homogeneous liquid. The derivation follows the standard
techniques of classical density functional theory as outlined in
\cite{Eva79} (similar approaches are also employed in
Ref.~\cite{dft0,dft1}).

    The derivation begins with the definition of the grand partition
function for $N$ particles at temperature $T$,
\be 
\Xi =Tr\{ e^{(H_N -\mu N)/k_B T} \}
\ee 
where
\be
Tr \equiv \frac{1}{h^{3N} N!}\sum_{N=0}^{\infty} \int d\vec{r}_1
d\vec{r}_2 ....d\vec{r}^N \int d\vec{p}_1 d\vec{p}_2 ....d\vec{p}^N
\ee
is the classical trace operator, with $\vec{r}_i$ and $\vec{p}_i$
being the position momentum of the $i^{th}$ atom, respectively,
$\mu$ is the chemical potential and $h$ denotes Plank's constant.
The N-body hamiltonian is denoted by $H_N=K+U+V$, where
\ben 
K &=&\sum_{i=1}^N \frac{p_i^2}{m_i} \nonumber \\
U &\equiv& U(\vec{r}_1,\cdots,\vec{r}_N ) \nonumber \\
V &=& \sum_{i=1}^{N} V_{\rm ext}(\vec{r}_i)
\een
Here $U$ denotes the interaction potential between particles in the
the system (including many body interactions), $K$ is the total
kinetic energy, with $m_i$ the mass of particles $i$ and $V_{\rm
ext}$ represents the interaction of atom $i$ with an external field.
The probability density for any particular phase space configuration
is given by
\be 
f_{\rm eq}= \Xi^{-1} e^{(H_N -\mu N)/k_B T}
\ee 

    The number density operator an N-body system is defined by by
\be
\hat{\rho}(\vec{r})=\sum_{i=1}^N \delta (\vec{r}-\vec{r}_i)
\ee
The equilibrium number density is obtained by averaging the density
operator with the equilibrium probability density,
\be
\rho(\vec{r})=\langle \hat{\rho}\rangle=Tr\{ \hat{\rho}f_{\rm eq} \}
\label{eq_density}
\ee
The PFC formalism will ultimately yield governing equations for the
evolution of the number density in Eq.~(\ref{eq_density}) on
diffusive time scales. It is noteworthy that the equilibrium
probability density, $f_{\rm eq}$, is a functional of
$\rho(\vec{r})$ \cite{Eva79}, an important property used in the
derivation below. This is because for a given $U$, $V_{\rm ext}$ is
uniquely determined by $\rho(\vec{r})$. Moreover, since $V_{\rm
ext}$ determines $f_{\rm eq}$, it follows that the equilibrium
density $f_{\rm eq}$ is a functional of $\rho(\vec{r})$. The details
of this argument are derived rigorously in in Ref.~\cite{Eva79} and
will not be reproduced here.

    The arguments of the previous paragraph also imply that for a given
$U$, the Helmholtz free energy, defined by
\be
{\cal F}[\rho]=Tr\{ f_{\rm eq} (K+U+k_B T \ln f_{\rm eq} ) \}
\label{free_eng_functional}
\ee
is also a functional of $\rho(\vec{r})$ \cite{Eva79}, as is also the
grand potential functional, defined by
\be
\Omega[\rho]=\int d\vec{r} \rho(\vec{r}) V_{\rm ext}(\vec{r})+ {\cal
F}[\rho] -\mu\int d\vec{r} \rho(\vec{r}) \label{grand_func}
\ee
It should be noted that the grand potential can be cast into a more
familiar form by substituting Eq.~(\ref{eq_density}) into
$\Omega[\rho]$ and exchanging the order of integration over
$\vec{r}$ and $Tr$ operation. Specifically, using the
straightforward results $\int d\vec{r} \rho(\vec{r}) V_{\rm
ext}(\vec{r})=-Tr\{f_{\rm eq} V \}$ and $-\mu \int d\vec{r}
\rho(\vec{r}) =-Tr\{f_{\rm eq} \mu N\}$, leads to the well known
result from statistical mechanics,
\be
\Omega[\rho]\equiv -k_B T \ln \Xi
\ee
The grand potential is important in that it can be used to relate
the chemical potential to the equilibrium density $\rho(\vec{r})$
according to $\delta \Omega[\rho]/\delta \rho(\vec{r})=0$ which
gives \cite{Eva79}
\be
\mu =V_{\rm ext}(\vec{r})+\frac{\delta {\cal F}[\rho]}{\delta
\rho(\vec{r})} \label{chem_pot}
\ee
Equation (\ref{chem_pot}) is fundamental to the theory of
non-uniform fluids in as it can, in principle, be used to calculate
the equilibrium density \cite{Eva79,dft1}.

    The properties of the free energy functional ${\cal F}[\rho]$ can be
be better elucidated by writing it as the sum of two terms
\be
{\cal F}[\rho]={\cal F}_0[\rho]-\Phi[\rho] \label{free_split}
\ee
where ${\cal F}_0$ represents the ideal case of non-interacting
particles, while $\Phi[\rho]$ represents the total potential energy
of interactions between the particles. Note that for a given $U$,
$\Phi$ is, once again, a functional of $\rho(\vec{r})$. Moreover,
for $U=0$ in Eq.~(\ref{free_eng_functional}) ${\cal F}_0[\rho]$
becomes
\be
{\cal F}_0[\rho]=k_B T\int d \vec{r} \rho(\vec{r}) \left( \ln(
\lambda \rho(\vec{r}))-1\right) \label{non_int_F}
\ee
where $\lambda=\sqrt{(h^2/2m\pi k_B T)}$ \cite{Eva79}.

    To further discuss the properties of ${\cal F}[\rho]$ for periodic phases,
it will be useful to expand the free energy in
Eq.~(\ref{free_split}) about the density, $\rho=\rho_l$, which
corresponds to the liquid side of the solid-liquid coexistence phase
diagram (at a given temperature). The change in free energy, ${\cal
F}_c \equiv {\cal F}[\rho]-{\cal F}[\rho_l]$, then becomes
\be
{\cal F}_c = \left({\cal F}_0[\rho]-{\cal F}_0[\rho_l]
\right)-\left( \Phi[\rho]-\Phi[\rho_l]\right) \label{free_split2}
\ee
The first term on the right hand side of Eq.~(\ref{free_split2}) can
be simplified by substituting $\rho=\rho_l+\delta \rho$ in the
non-logarithmic expressions of ${\cal F}_0[\rho]$, giving
\be
( {\cal F}_0[\rho]-{\cal F}_0[\rho_l] )=k_B T \int d\vec{r}
(\rho \ln(\rho/\rho_l) - \delta \rho) \label{F0}
\ee
In arriving at Eq.~(\ref{F0}), use was made of the property $\int d
\vec{r} \delta \rho=0$ in the periodic state. The interaction term,
$\left( \Phi[\rho]-\Phi[\rho_l]\right)$, can also be expanded
functionally in $\delta \rho(\vec{r})$ about $\rho_l$ by defining
\ben
C_1(\vec{r}) &\equiv & \frac{\delta \Phi[\rho(\vec{r})]}{\delta
\rho(\vec{r})} \nline C_2(\vec{r}_1,\vec{r}_2) &\equiv&
\frac{\delta^2 \Phi}{\delta \rho(\vec{r}_1) \delta \rho(\vec{r}_2)}
\nline C_3(\vec{r}_1,\vec{r}_2,\vec{r}_3) &\equiv& \frac{\delta^3
\Phi}{\delta \rho(\vec{r}_3) \delta \rho(\vec{r}_1) \delta
\rho(\vec{r}_2)}  \nline && \cdots \label{ap:C_funcs}
\een

Using Eq.~(\ref{F0}) and (\ref{ap:C_funcs}) in
Eq.~(\ref{free_split2}) finally gives
\ben
\frac{{\cal F}_c}{k_B T}&=&\int d\vec{x} \left[ \rho(\vec{r}) \ln
\left( \frac{\rho(\vec{r})}{\rho_l}\right)
-\delta \rho(\vec{r}) \right] \nonumber \\
-\frac{1}{2} &\int& d\vec{r}_1 d\vec{r}_2 \delta \rho(\vec{r}_1)
C_{2}(\vec{r}_1, \vec{r}_2) \delta \rho(\vec{r}_2) + \nonumber \\
-\frac{1}{6} &\int& d\vec{r}_1 d\vec{r}_2 d\vec{r}_3 \delta
\rho(\vec{r}_1) C_{3}(\vec{r}_1, \vec{r}_2, \vec{r}_3) \delta
\rho(\vec{r}_2) \delta \rho(\vec{r}_3) \nonumber \\ + &\cdots &
\label{ap:free_eng2}
\een
The function $C_{2}$ is the two point direct correlation function of
an isotropic fluid and it is usually denoted $C_{ij} \equiv
C_{2}(r_{12})$, where $r_{12}\equiv |\vec{r}_1-\vec{r}_2|$. The
function $C_{3}$ is the three point correlation function, etc.

\section{Simple Binary Alloy Model}
\label{app:simple}

    This appendix goes through the expansion required to arrive at the
simplified alloy model presented in section \ref{sect:alloys}. The
starting point is the definition of the following three fields,
\ben  
n&=&(\rho-\bar{\rho})/\bar{\rho} \nline
n_A&=&(\rho_A-\bar{\rho}_A)/\bar{\rho} \nline
n_B&=&(\rho_B-\bar{\rho}_B)/\bar{\rho} \nline
\een 
where overbars denote averages. The field $n$ can equivalently be
written as
\be 
n=n_A + n_B = (\rho_A+\rho_B)/\bar{\rho} -
(\bar{\rho}_A+\bar{\rho}_B)/\bar{\rho}
\ee 
Furthermore, the definitions of $c=\rho_A/\rho$ and $\delta c=1/2-c$
(see Section \ref{sect:alloys}) can be used to re-write
\ben
\delta c &=& \frac{\rho_B-\rho_A}{2\rho}
 = \frac{\bar{\rho}(n_B-n_A)+\bar{\rho}_B-\bar{\rho}_A}{2\rho},
\een 
which can further be used to define a new field
\be 
\delta N \equiv 2\rho\delta c/\bar{\rho}
 = (n_B-n_A)+\frac{\bar{\rho}_B-\bar{\rho}_A}{\bar{\rho}}.
\ee 

    Next, the fields $\rho$, $\delta \rho=\rho-\rho_l$, $c$ and $\delta
c$ in Eq.~(\ref{eq:freeb}) are expressed in terms of $n$ and $\delta
N$. Following that the free energy is expanded with respect to $n$
and $\delta N$ up to order four (noting that terms of order $n$ or
$\delta N$ can be dropped since they integrate to zero in the free
energy functional as they are all defined around their average
values). Carrying out these expansions gives,
\ben
\frac{{\cal F}}{\bar{\rho}k_B T} &=& \int d\vec{r} \Big[
f_o+\frac{n}{2}\left(1-\bar{\rho}\frac{C_{AA}+C_{BB}+2C_{AB}}{4}\right)n
\nline && -\frac{n^3}{6} +
\frac{n^4}{12}+\frac{\beta}{2\bar{\rho}}(1- n+n^2-n^3)\delta N
\nline &&
+\frac{C_{AA}-C_{BB}}{4\bar{\rho}}\Big((\bar{\rho}-\rho_{\ell})^2
+(\bar{\rho}^2-\rho_{\ell}^2)n \nline && + \rho_{\ell}^2(n^2-n^3)
\Big)\delta N \nline &&+\frac{\delta N}{2} \left(1-
\frac{C_{AA}+C_{BB}-2C_{AB}}{4}\right)\delta N \nline
&&-(n-n^2)\frac{\delta N^2}{2}+\frac{\delta N^4}{12} \Big]
\een
where
\ben 
f_o &=& \ln\left(\frac{\bar{\rho}}{2\rho_{\ell}}\right)
-(1-\rho_{\ell}/\bar{\rho}) -\bar{\rho}C_{AB}^0/4 \nline &&
-\frac{1}{8}(
\bar{\rho}+2\rho_{\ell}^2/\bar{\rho}-4\rho_{\ell})(C_{AA}^0+C_{BB}^0)
\nline
\een
Simplifying further, paying particular attention to the $n^2$ terms
and substituting the explicit forms for $C_{i,j}^n$ gives,
\ben
\frac{{\cal F}}{\bar{\rho}k_B T} &=& \int d\vec{r} \Big[ f_o+
B^{\ell}\frac{n^2}{2} -\frac{n^3}{6} +
\frac{n^4}{12}+nF\nabla^2n+nG\nabla^4n\nline &&
+\frac{\beta}{2\bar{\rho}}(1-n-n^3)\delta N \nline &&
+\frac{C_{AA}-C_{BB}}{4\bar{\rho}}\Big((\bar{\rho}-\rho_{\ell})^2
+(\bar{\rho}^2-\rho_{\ell}^2)n \nline && - \rho_{\ell}^2 n^3)
\Big)\delta N \nline &&+\frac{\delta N}{2} \left(1-
\frac{C_{AA}+C_{BB}-2C_{AB}}{4}\right)\delta N \nline
&&-\frac{\delta N^2}{2}n+\frac{\delta N^4}{12} \label{Eq:this} \Big]
\een
where
\ben
B^{\ell} &=& 1- \bar{\rho}(C^o_{AA}+C^o_{BB}+2C^o_{AB})/4 \nline
&&+\left(\frac{\beta}{\bar{\rho}}
+\frac{\rho_{\ell}^2}{2\bar{\rho}}(C^o_{AA}-C^o_{BB})\right) \delta
N +\delta N^2\nline F &=& -
\bar{\rho}(C^2_{AA}+C^2_{BB}+2C^2_{AB})/4 \nline
&&+\frac{\rho_{\ell}^2}{2\bar{\rho}}(C^2_{AA}-C^2_{BB})\,\delta N
\nline G &=& - \bar{\rho}(C^4_{AA}+C^4_{BB}+2C^4_{AB})/4 \nline
&&+\frac{\rho_{\ell}^2}{2\bar{\rho}}(C^4_{AA}-C^4_{BB})\,\delta N
\nline
\een

    In what follows it is assumed that $\delta N$ varies on length
scales much larger than $n$. This is a reasonable on long time
(diffusion) times scales, where solute and host atoms intermix on
length scales many times larger than the atomic radius. This
assumption allows terms of order $n$ to be eliminated from the free
energy, i.e.,
\ben
\frac{{\cal F}}{\bar{\rho}k_B T} &=& \int d\vec{r} \Big[ f_o+
B^{\ell}\frac{n^2}{2} -\frac{n^3}{6} +
\frac{n^4}{12}+nF\nabla^2n+nG\nabla^4n\nline &&+\frac{\delta N}{2}
\left(1- \frac{C_{AA}+C_{BB}-2C_{AB}}{4}\right)\delta N
+\frac{\delta N^4}{12} \nline &&
+\frac{\beta}{2\bar{\rho}}(1-n^3)\delta N \nline &&
+\frac{C_{AA}-C_{BB}}{4\bar{\rho}}\Big((\bar{\rho}-\rho_{\ell})^2
 - \rho_{\ell}^2n^3)
\Big)\delta N  \Big]
\een
The previous equation can finally be cast into a form similar to
that presented in Section \ref{sec:simple} of the text,
\ben 
\frac{{\cal F}}{\bar{\rho}k_B T} &=& \int d\vec{r} \Big[
f_o+\frac{n}{2} \left[ B^{\ell} +B^s
(2R^2\nabla^2+R^4\nabla^4)\right] n \nline &-&\frac{n^3}{6} +
\frac{n^4}{12}+\frac{w}{2}\delta N^2 +\frac{\delta N^4}{12} +
\frac{L^2}{2} |\nabla \delta N|^2 \nline &+& \gamma \delta N
+\frac{H^4}{2}\delta N \nabla^4 \delta N \Big] \nline
\een
where
\ben 
B^s &=& F^2/(2G) \nline R &=& \sqrt{2G/F} \nline w &=&
(1-(C^0_{AA}+C^0_{BB}-2C^0_{AB})/2) \nline L^2 &=&
(C^2_{AA}+C^2_{BB}-2C^2_{AB})/2) \nline H^2 &=&
-(C^4_{AA}+C^4_{BB}-2C^4_{AB})/2)\nline \gamma &=&
\frac{\beta}{2\bar{\rho}}(1-n^3) \nline &+&
\frac{C_{AA}-C_{BB}}{4\bar{\rho}}\Big((\bar{\rho}-\rho_{\ell})^2
 - \rho_{\ell}^2n^3\Big)
\label{app:coeffs}
\een

    The dependence of the coefficients in
$B^l$, $B^{\ell}$ and $R$ on the density difference can be
explicitly obtained by expanding them in $\delta N$ as well. This
gives,
\ben
B^{\ell} &=& B^{\ell}_0+B^{\ell}_1\,\delta N + B^{\ell}_2\, \delta
N^2 \nline B^s &=& B^s_0+B^s_1\,\delta N + B^s_2\, \delta N^2+\cdots
\nline R &=& R_0+R_1\,\delta N + R_2\, \delta N^2+\cdots
\label{BR_exp}
\een 
where
\ben 
B^{\ell}_0 &=& 1- \bar{\rho}\,\cb{0} \nline B^{\ell}_1 &=&
\frac{\beta}{\bar{\rho}} +\frac{\rho_{\ell}^2}{2\bar{\rho}}\dChat{0}
\nline B^{\ell}_2 &=& 1
\een
and
\ben 
\cb{n} &\equiv& (C^{(n)}_{AA}+C^{(n)}_{BB}+2C^{(n)}_{AB})/4 \nline
\dChat{n} &\equiv& C^{(n)}_{AA}-C^{(n)}_{BB},
\een
while
\ben
B^s_0 &=& - \bar{\rho}\frac{(\cb{2})^2}{\cb{4}} \nline B^s_1 &=&
-\frac{\cb{2}\rho_{\ell}^2}{4\bar{\rho}\cb{4}^2}
(\cb{2}\dChat{4}-2\cb{4}\dChat{2}) \nline B^s_2 &=&
-\frac{\rho_{\ell}^2}{8\bar{\rho}^3\cb{4}^3}
(\cb{2}\dChat{4}-\cb{4}\dChat{2})^2
\een 
and
\ben 
R_0 &=& \sqrt{\frac{2\cb{4}}{\cb{2}}} \nline R_1/R_0 &=&
-\frac{\rho_{\ell}^2}{4\bar{\rho}^2\cb{4}\cb{2}}
(\cb{2}\dChat{4}-\cb{4}\dChat{2}) \nline R_2/R_0 &=&
-\frac{\rho_{\ell}^4}{32\bar{\rho}^4\cb{4}^2\cb{2}^2}
(\cb{2}\dChat{4}-\cb{4}\dChat{2}) \times \nline
&&(\cb{2}\dChat{4}+3\cb{4}\dChat{2})
\een 

\end{document}